\documentclass[]{jfm}
\usepackage{natbib} 
\usepackage[breaklinks,colorlinks=true,allcolors=blue]{hyperref}
\usepackage{overpic} 
\usepackage{booktabs} 
\usepackage{dcolumn}
\usepackage{color,amsmath,amssymb,bm}
\usepackage{mathabx}
\usepackage{mathtools}
\usepackage{empheq}
\usepackage{physics} 
\usepackage{fancyhdr}
\newcommand{\lk}{\left(}
\newcommand{\rk}{\right)}
\newcommand{\pa}{\partial}
\renewcommand{\d}{{\mathrm{d}}}
 
\newcommand{\fa}[1]{{\textit{#1}}} 
\DeclareMathAlphabet{\mathsfbi}{OT1}{\sfdefault}{bx}{sl}
\DeclareMathVersion{sfletters}
\SetSymbolFont{letters}{sfletters}{OML}{ntxsfmi}{b}{it}

\makeatletter
\newcommand{\mathbfsbilow}[1]{%
  \text{\mathversion{sfletters}$\m@th#1$}%
}

\DeclareRobustCommand{\tensor}[1]{%
  \begingroup
  \ifcat\noexpand #1\relax
    \edef\greek@test{\detokenize{#1}}%
    \edef\greek@test{\expandafter\@cdr\greek@test\@nil}%
    \edef\greek@test{\expandafter\@car\greek@test\@nil}%
    \edef\x{\the\lccode\expandafter`\greek@test}%
    \edef\y{\number\expandafter`\greek@test}%
    \ifnum\x=\y\relax
      \mathbfsbilow{#1}%
    \else
      \mathsfbi{#1}%
    \fi
  \else
    \mathsfbi{#1}%
  \fi
  \endgroup
}

\title{Attenuation mechanism of wall-bounded turbulence by heavy finite-size particles}

\author{
        Yutaro Motoori
        \corresp{\email{y.motoori.es@osaka-u.ac.jp}}
		\and 
        Susumu Goto
        \corresp{\email{s.goto.es@osaka-u.ac.jp}}
        }

\affiliation{Graduate School of Engineering Science, The University of Osaka, 1-3
Machikaneyama, Toyonaka, Osaka, 560-8531, Japan}



\begin{document}
\maketitle

\begin{abstract}
To elucidate the attenuation mechanism of wall-bounded turbulence due to heavy
small particles, we conduct direct numerical simulations (DNS) of turbulent
channel flow laden with finite-size solid particles. When particles cannot
follow the swirling motions of wall-attached vortices, vortex rings are
created around the particles. These particle-induced vortices lead to
additional energy dissipation, reducing the turbulent energy production from the
mean flow. This mechanism results in the attenuation of turbulent kinetic energy,
which is more significant when the Stokes number of particles is larger or
particle size is smaller under the condition that the volume fraction of
particles is fixed. Moreover, we propose the method to quantitatively predict
the degree of turbulence attenuation without using DNS data by estimating the
additional energy dissipation rate in terms of particle properties.
\end{abstract}


\section{Introduction \label{sec:intro}}

The addition of small particles can attenuate turbulence. This phenomenon has
been demonstrated by many experiments since the last century. For example,
\citet{Gore1989} compiled experimental results on turbulence modulation,
demonstrating that when the particle diameter is about $0.1$ times smaller than
the integral length of turbulence, the particles can attenuate turbulence;
otherwise, the larger particles enhance it. Although many authors
\citep{Hosokawa2004, Righetti2004, Tanaka2008, Noguchi2009, Yu2021} proposed
other parameters that characterize turbulence modulation, there is a consensus,
supported by experiments \citep{Maeda1980, Tsuji1984, Kulick1994, Rogers1991,
Fessler1999, Kussin2002, Yang2005} and direct numerical simulations (DNS)
\citep{Ferrante2003, TenCate2004, Vreman2007, Zhao2010, Abdelsamie2012,
Zhao2013, Li2016, Liu2017, Mortimer2019, Oka2022, Peng2023}, that small heavy
particles can lead to turbulence attenuation.

In addition to these studies, to accumulate a significant body of knowledge on
the turbulence modulation due to particles, many researchers \citep{TenCate2004,
Burton2005, Lucci2010, Yeo2010, Bellani2012, Wang2014, Schneiders2017,
Uhlmann2017, Oka2022, Shen2022, Peng2023} conducted DNS of turbulence
interacting with finite-size spherical particles in a periodic box. Among these
studies, it is crucial to highlight the observation reported by
\citet{TenCate2004} that the energy dissipation rate produced by the relative
motion between particles and fluid is a key quantity for significant attenuation
of turbulent kinetic energy. In addition to their study, many authors
\citep{Squires1990, Elghobashi1994, Kulick1994, Paris2001, Hwang2006, Mando2009,
Balachandar2010, Yeo2010, Wang2014, Peng2019b} also emphasized the importance of
the additional energy dissipation rate for turbulence attenuation. Recently, our
group \citep{Oka2022} conducted DNS of periodic turbulence laden with
finite-size particles and derived a formula to describe the turbulence
attenuation rate. This formula is based on the physical picture of the
turbulence attenuation that the additional energy dissipation due to particles
bypasses the energy cascade. \citet{Balachandar2024} further developed this view
to model the energy balance between particles and turbulence at the
subgrid scales. Thus, the additional energy dissipation due to particles is
crucial for turbulence attenuation. However, most of these results were obtained
through numerical analyses of periodic turbulence without walls, although the
pioneering studies on turbulence modulation by particles were made by
experiments of air turbulence bounded by solid walls \citep{Tsuji1982,
Tsuji1984, Kulick1994}. Therefore, the next important issue is understanding the
modulation mechanism of wall-bounded turbulence. This is the target of the
present study.

\citet{Kulick1994} experimentally demonstrated for turbulent channel flow that
heavy copper particles attenuated turbulence intensity more effectively than
glass particles. They concluded that as the relaxation time $\tau_p$ of
particles becomes longer than the time scale $\tau_f$ of fluid motion, i.e., as
the Stokes number $St=\tau_p/\tau_f$ increases, the degree of turbulence
attenuation becomes more significant. Note that the relaxation time is longer
for heavier particles if they have the same diameter. Similarly to turbulence in
a periodic box, the attenuation of wall-bounded turbulence for large $St$ is
also explained by a relative velocity between particles and fluid because it
leads to significant energy dissipation around particles \citep{Paris2001,
Peng2019a}. To understand the modulation mechanism of wall-bounded turbulence,
many DNS studies were conducted for turbulent channel flow laden with heavy
pointwise particles \citep{Dritselis2008, Dritselis2011, Lee2015, Wang2019,
Zhou2020a} and finite-size ones \citep{Kajishima2001, Uhlmann2008, Zeng2008,
Shao2012, Fornari2016, Li2016, Yu2017, Peng2019a, Peng2019b, Costa2020, Yu2021,
Xia2021, Muramulla2020, Costa2021, Brandt2022, Shen2024}. For example,
\citet{Zhao2013} and \citet{Yu2021} discussed interactions between particles and
turbulence on the basis of turbulent kinetic energy and Reynolds stress budgets.
\citet{Peng2019b} depicted the schematic of turbulence modulation in the viscous
sublayer and the buffer layer. Recently, \citet{Shen2024} investigated particle
parameter dependence of turbulence modulation especially in the near-wall
region. However, a framework for predicting the degree of attenuation is
incomplete because we do not fully understand the concrete picture of the
attenuation mechanism in wall-bounded turbulence.

The purposes of the present study are (I) to understand the attenuation
mechanism of wall-bounded turbulence by small heavy particles and (II) to
predict the degree of turbulence attenuation in terms of particle properties.
For these purposes, we conduct DNS of turbulent channel flow with finite-size
spherical particles, systematically changing the particle diameter, mass density
and turbulence Reynolds number. In particular, we focus on heavy particles as
small as the buffer-layer coherent structures. To investigate the interaction
between particles and turbulence, we assume that the gravitational effects are
negligible. 

In the following, we first describe the coupled DNS method for turbulent channel
flow with particles (\S\:\ref{sec:method}). Then, we examine the modulation of
coherent vortices in real space (\S\:\ref{sec:vortex}) and quantitatively
investigate the modulation of energy transfer mechanism
(\S\S\:\ref{sec:energy}--\ref{sec:mechanism}). Then, based on this attenuation
mechanism, we develop an argument to predict the degree of turbulence
attenuation (\S\:\ref{sec:discussion}).

\section{Methods\label{sec:method}}
\subsection{Direct numerical simulations}

We numerically simulate turbulence laden with finite-size particles between two
parallel planes. Under the non-slip boundary conditions on the surfaces of the
particles and walls, the flow obeys the Navier--Stokes equation,
\begin{align}
\label{eq:NS}
\frac{\pa\vb*{u}}{\pa t}
+
(\vb*{u}\cdot\vb*{\nabla})\vb*{u}
=
\vb*{\nabla}\cdot \tensor{T}
+
\vb*{f}^{ex}
,  
\end{align}
and the continuity equation,
\begin{align}
\vb*{\nabla}
\cdot
\vb*{u}
=
0 ,
\end{align}
where $\vb*{u}(\vb*{x},t)$ and $p(\vb*{x},t)$ are the fluid velocity and the
pressure at position $\vb*{x}$ and time $t$, respectively. In (\ref{eq:NS}),
$\tensor{T}$ is the stress tensor divided by the mass density for a Newtonian
fluid, and $\vb*{f}^{ex}$ ($=(f^{ex}_x, 0, 0)$) is the force due to a constant
pressure difference. We solve these equations using the immersed boundary method
(IBM) proposed by \citet{Breugem2012}. In this method, similarly to the original
method by \citet{Uhlmann2005}, we introduce a force between particles and fluid.
More concretely, instead of solving (\ref{eq:NS}) under the boundary conditions
on the particle surfaces, we solve
\begin{align}
\label{eq:NS-IBM}
\frac{\pa\vb*{u}}{\pa t}
+
(\vb*{u}\cdot\vb*{\nabla})\vb*{u}
=
-
\frac{1}{\rho_f}\vb*{\nabla}p
+
\nu\nabla^2\vb*{u}
+
\vb*{f}^{ex}
+
\vb*{f}^{{IBM}}
\end{align}
in the entire computational domain including the inside of particles. Here,
$\rho_f$ and $\nu$ denote the fluid mass density and kinematic viscosity,
respectively, and $\vb*{f} ^{{IBM}}$ is the force per unit mass to satisfy the
non-slip boundary condition at the Lagrangian points distributed on the surface
of each particle.

\begin{table}
\centering
	\caption{
		Particle parameters. 
    \label{table:parameter}}
\vspace{2mm}
\begin{tabular}{D{.}{.}{1}D{.}{.}{2}D{.}{.}{2}D{.}{.}{2}D{.}{.}{1}D{.}{.}{2}D{.}{.}{3}D{.}{.}{2}}
\addlinespace[2mm]
\multicolumn{1}{p{0.5cm}}{$Re_\tau$} &
\multicolumn{1}{p{1.1cm}}{\,\,\,\,\,\,$D^+$} &
\multicolumn{1}{p{0.2cm}}{\,$D/h$} &
\multicolumn{1}{p{1.3cm}}{\,\,\,\,\,\,\,$D/\Delta$} &
\multicolumn{1}{p{0.9cm}}{\,$\rho_p/\rho_f$} & 
\multicolumn{1}{p{0.9cm}}{\,\,\,\,$St_+$} & 
\multicolumn{1}{p{0.6cm}}{\,\,$St_h$} & 
\multicolumn{1}{p{1.2cm}}{\,\,\,\,$N_p$}
\\ 
\addlinespace[0.2mm]
\hline
\addlinespace[0.2mm]
$512$ & $16$ & 0.031 & $8$          & $2$      & $28$     & 0.056 & $8192$ \\
$512$ & $16$ & 0.031 & $8$          & $8$      & $114$    & 0.22  & $8192$ \\
$512$ & $16$ & 0.031 & $8$          & $32$     & $455$    & 0.89  & $8192$ \\
$512$ & $16$ & 0.031 & $8$          & $128$    & $1820$   & 3.6   & $8192$ \\
\addlinespace[1mm]
$512$ & $32$ & 0.063 & $16$         & $2$      & $114$    & 0.22  & $1024$ \\
$512$ & $32$ & 0.063 & $16$         & $8$      & $455$    & 0.89  & $1024$ \\
$512$ & $32$ & 0.063 & $16$         & $32$     & $1820$   & 3.6   & $1024$ \\
$512$ & $32$ & 0.063 & $16$         & $128$    & $7282$   & 14    & $1024$ \\
\addlinespace[1mm]
$512$ & $64$ & 0.13  & $32$         & $2$      & $455$    & 0.89  & $128$  \\
$512$ & $64$ & 0.13  & $32$         & $8$      & $1820$   & 3.6   & $128$  \\
$512$ & $64$ & 0.13  & $32$         & $32$     & $7282$   & 14    & $128$  \\
$512$ & $64$ & 0.13  & $32$         & $128$    & $29127$  & 57    & $128$  \\
%
\addlinespace[4mm]
$360$ & $16$ & 0.044 & $8$          & $2$      & $28$     & 0.078  & $2847$ \\
$360$ & $16$ & 0.044 & $8$          & $8$      & $114$    & 0.32  & $2847$ \\
$360$ & $16$ & 0.044 & $8$          & $32$     & $455$    & 1.3   & $2847$ \\
$360$ & $16$ & 0.044 & $8$          & $128$    & $1820$   & 5.1   & $2847$ \\
\addlinespace[1mm]
$360$ & $32$ & 0.089 & $16$         & $2$      & $114$    & 0.32  & $356$ \\
$360$ & $32$ & 0.089 & $16$         & $8$      & $455$    & 1.3   & $356$ \\
$360$ & $32$ & 0.089 & $16$         & $32$     & $1820$   & 5.1   & $356$ \\
$360$ & $32$ & 0.089 & $16$         & $128$    & $7282$   & 20    & $356$ \\
%
\addlinespace[4mm]
$256$ & $16$ & 0.063 & $8$          & $2$      & $28$     & 0.11  & $1024$ \\
$256$ & $16$ & 0.063 & $8$          & $8$      & $114$    & 0.44  & $1024$ \\
$256$ & $16$ & 0.063 & $8$          & $32$     & $455$    & 1.8   & $1024$ \\
$256$ & $16$ & 0.063 & $8$          & $128$    & $1820$   & 7.1   & $1024$ \\
\addlinespace[1mm]
$256$ & $32$ & 0.13  & $16$         & $2$      & $114$    & 0.44  & $128$ \\
$256$ & $32$ & 0.13  & $16$         & $8$      & $455$    & 1.8   & $128$ \\
$256$ & $32$ & 0.13  & $16$         & $32$     & $1820$   & 7.1   & $128$ \\
$256$ & $32$ & 0.13  & $16$         & $128$    & $7282$   & 28    & $128$ \\
%
\addlinespace[4mm]
$180$ & $16$ & 0.089 & $8$          & $2$      & $28$     & 0.16  & $356$ \\
$180$ & $16$ & 0.089 & $8$          & $8$      & $114$    & 0.63  & $356$ \\
$180$ & $16$ & 0.089 & $8$          & $32$     & $455$    & 2.5   & $356$ \\
$180$ & $16$ & 0.089 & $8$          & $128$    & $1820$   & 10   & $356$ \\
\vspace{1mm}
\end{tabular}

\end{table}

We consider the motion of spherical solid particles with a mass density
$\rho_p$, volume $V_p$ and moment of inertia $I_p$. The equations of motion of a
particle with velocity $\vb*{u}_p(t) = \d \vb*{x}_p/\d t$ and angular velocity
$\vb*{\omega}_p(t)$ are 
\begin{align}
\label{eq:p1}
\rho_p V_p
\frac{\d \vb*{u}_p}{\d t}
&=
\oint\nolimits_{\pa V_p}\: \tensor{T} \cdot
\vb*{n}^{\text{\textcircled{$p$}}}\:\d S
+
\rho_f V_p \vb*{f}^{ex}
+
\vb*{F}^{\leftrightarrow {p}}
\end{align}
and
\begin{align}
\label{eq:p2}
I_p 
\frac{\d \vb*{\omega}_p}{\d t}
&=
\oint\nolimits_{\pa V_p}\: \vb*{r}\times (\tensor{T}
\cdot\vb*{n}^{\text{\textcircled{$p$}}})\:\d S
+
\vb*{T}^{\leftrightarrow {p}} .
\end{align}
Here, $\vb*{n}^{\text{\textcircled{$p$}}}$ denotes the outward-pointing normal
vector on the surface $\pa V_p$ of the particle, $\vb*{r}$ is the position
vector relative to the particle centre, and $\vb*{F}^{\leftrightarrow {p}}$ and
$\vb*{T}^{\leftrightarrow {p}}$ are the force and torque acting on the particle
due to collisions with other particles or solid walls, respectively. In the IBM,
we evaluate the first term of the right-hand side of (\ref{eq:p1}) with
\begin{align}
\oint\nolimits_{\pa V_p}\: \tensor{T} \cdot \vb*{n}^{\text{\textcircled{$p$}}}\:\d S
=
-
\rho_f
\sum_{\ell = 1}^{N_\ell}
\vb*{f}^{{IBM}}_{\ell} \: \Delta V_{\ell}
+
\rho_f\frac{\d}{\d t}\int_{V_p}\:\vb*{u}\:\d V 
-
\rho_f V_p \vb*{f}^{ex} . 
\label{eq:tauf}
\end{align}
Here, $\vb*{f}^{{IBM}}_{\ell}$ is the fluid force (per unit mass) by particles
acting at the $N_\ell$ Lagrangian points distributed on the particle surface,
and $\Delta V_\ell$ is the volume of a Lagrangian grid cell. The first term of
the right-hand side of (\ref{eq:p2}) is given by $\vb*{r}\times$(\ref{eq:tauf}).
Then, we evaluate the last terms ($\vb*{F}^{\leftrightarrow {p}}$ and
$\vb*{T}^{\leftrightarrow {p}}$) using an elastic model \citep{Glowinski2001}
for the normal component of the contact force. We neglect the frictional force
and collision torque.  

To couple flow and particle motion, we alternately integrate their governing
equations with the second-order Crank--Nicolson method for the viscous term of
the fluid and explicit third-order low-storage three-step Runge--Kutta method
for the others. Details of the numerical method are as described in
\citet{Breugem2012}.

\subsection{Parameters\label{sec:parameter}}

We simulate turbulent channel flow at four values of the friction Reynolds
number: $Re_\tau=u_\tau h / \nu = 512$, $360$, $256$ and $180$, where $u_\tau$
is the friction velocity and $h$ is the channel half-width. Note that since the
external driving force $f_x^{ex}$ is constant, the frictional force acting on
the wall, and therefore $Re_\tau$ are kept constant even in the case with
particles. The computational domain sides are $4h$, $2h$ and $2h$ in the
streamwise $x$, wall-normal $y$ and spanwise $z$ directions, respectively. The
grid width is $\Delta^+=2$ for all directions. Here, ${\:\cdot\:}^+$ denotes a
quantity normalised by $u_\tau$ and $\nu$. 

We add particles into turbulence in a statistically steady state. The particles
are characterised by three parameters related to the diameter $D$, mass density
$\rho_p$ and the number $N_p$ of particles. Fixing the volume fraction
$\Lambda_0$ ($=N_p V_p/ (16h^3)$) at $0.0082$, we change the other two
parameters: $D$ and $\rho_p$. The particle diameters are $D^+=16$, $32$ and $64$
for $Re_\tau=512$; $D^+=16$ and $32$ for $Re_\tau=360$ and $256$; and $D^+=16$
for $Re_\tau = 180$. We list in table~\ref{table:parameter} the parameters of
particles. The particle diameters are comparable to or a few times larger than
the diameter of tubular quasi-streamwise vortices in the buffer layer; but they
are sufficiently smaller than the channel-half width (i.e.~$D/h \lesssim 0.1$).
All particles are resolved to satisfy $D/\Delta \geq 8$, which is the same as in
our previous studies \citep{Oka2022, Awai2025}. For the mass density $\rho_p$,
we set it to be larger than that $\rho_f$ of the fluid: $\rho_p/\rho_f=2$, $8$,
$32$ and $128$. We neglect gravitational effect. In this situation, we observe
only turbulence attenuation because the relative velocity between particles and
fluid is not enough to enhance turbulence. This is consistent with an indicator
for turbulence enhancement proposed by \citet{Yu2021}; see
Appendix~\ref{sec:enhancement}. In addition, we also recently discuss turbulence
enhancement in the channel flow with gravity \citep{Motoori2025a}.

\begin{figure}
	\centering
	\includegraphics[width=90mm]{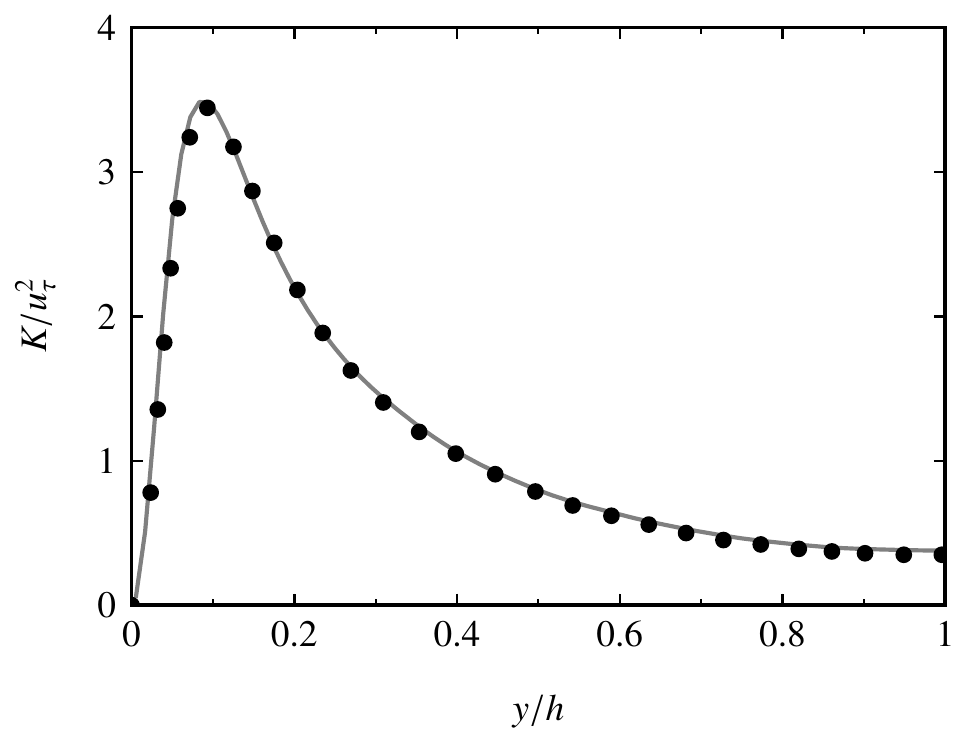}
	\caption{
	Wall-normal profiles of the mean turbulent kinetic energy $K(y)$. The
	circles show the result by \citet{Yu2021}. The parameters are $D/h=0.1$,
	$\rho_p/\rho_f=2$, $\Lambda_0=0.0236$, $u_g/U_b=0.159$ and $U_b (2h) /
	\nu=5746$. Data are extracted from figure~8(\fa{a}) of their paper. The line
	shows the present DNS result with the same parameters. 
	\label{fig:Kvalidation}}
\end{figure}

In the following, we show results on turbulence modulation using the relaxation
time $\tau_p$ of particles to the fluid motion instead of $\rho_p/\rho_f$. Here,
we define the relaxation time as $\tau_p = {\rho_p D^2}/{(18 \rho_f \nu)}$ by
assuming the Stokes drag. We list in table~\ref{table:parameter} the values of
$St_+=\tau_p/\tau^+$, defined with the wall friction time $\tau^+$ ($=\nu /
u_\tau^2$), and $St_h = \tau_p/\tau_h$, defined with the largest eddy turnover
time $\tau_h = h/u_\tau$. The examined particles have relaxation time $\tau_p$
in a wide range between the swirling time scales of the smallest and largest
vortices. Specifically, the shortest relaxation time ($St_+=28$) is comparable
to the time scale of coherent structures in the buffer layer
\citep{Soldati2009}. In other words, none of the simulated particles can
thoroughly follow the swirl of streamwise vortices in the buffer layer. On the
other hand, $St_h$ ($= St_+ Re_\tau$) represents particles' ability to follow
the largest vortices, that is, the wall-attached vortices in the outer layer.
Particles with $St_h \lesssim 1$ can follow the swirling motions of the
outer-layer vortices, whereas those with $St_h \gtrsim 1$ cannot follow them. 

\subsection{Validation}

Before showing results in the next section, we validate our DNS by simulating
particle-laden turbulence in the same system as investigated by \citet{Yu2021}.
For this, we impose a time-dependent pressure gradient to maintain a constant
flow rate and apply the gravitational force to ensure that the ratio between the
characteristic settling velocity $u_g$ ($=\sqrt{4D(\rho_p/\rho_f-1)g/3}$ with
$g$ being the magnitude of the gravitational acceleration) of particles and the
mean bulk velocity $U_b$ is $0.159$. Figure~\ref{fig:Kvalidation} shows the
wall-normal profiles of the mean turbulent kinetic energy,
\begin{align}
\label{eq:K}
K(y)
=
\frac{1}{2} \overline{ |{\vb*{u}^\prime}(\vb*{x},t)|^2 } ,
\end{align} 
in the statically steady state. Here, $\overline{\:\cdot\:}$ denotes the average
in the streamwise and spanwise directions and time over the fluid region outside
particles, and $\vb*{u}^\prime(\vb*{x},t)$ ($= \vb*{u}(\vb*{x},t) - \vb*{U}(y)$)
is the fluctuation fluid velocity, where $\vb*{U}(y)$
($=\overline{\vb*{u}(\vb*{x},t)}$) is the mean velocity. We can confirm that our
results are in good agreement with those of \citet{Yu2021}. In the following
analyses, we discuss the statistically steady state after adding the particles.

\section{Attenuation mechanism}

\subsection{Modulation of coherent vortices\label{sec:vortex}}

\begin{figure}
	\centering
	\includegraphics[width=\textwidth]{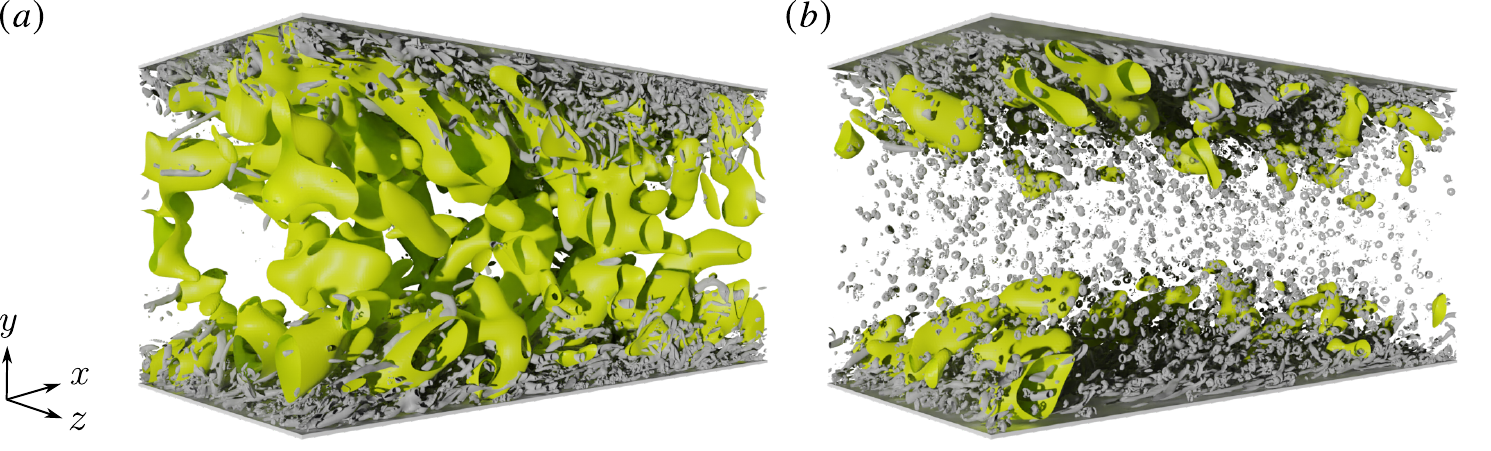}
	\caption{
	Visualisation of coherent vortices in turbulence at $Re_\tau = 512$ laden
	with particles with the same diameter ($D/h=0.031$, i.e.~$D^+=16$) but
	different values of the Stokes number: (\fa{a}) $St_h=0.056$ and (\fa{b})
	$3.6$. Grey vortices are identified by positive isosurfaces of the second
	invariant $Q$ of the velocity gradient tensor. Yellow vortices are
	identified by the second invariant
	$\mathrlap{\widetilde{\phantom{W}}}\,Q^{(\ell)}$ of the velocity gradient
	tensor coarse-grained at $\ell = 0.2h$. We set the thresholds as
	$Q^+=7.0\times 10^{-3}$ and
	$\mathrlap{\widetilde{\phantom{W}}}\,Q^{(\ell)+}=3.0\times 10^{-5}$. 
    \label{fig:vis}}
\end{figure}

First, let us examine the modulation of coherent vortices in turbulence.
Figure~\ref{fig:vis} shows vortices in turbulence at $Re_\tau=512$ laden with
the smallest ($D/h=0.031$, i.e.~$D^+=16$) particles. The Stokes number differs
between the panels: (\fa{a}) $St_h=0.056$ and (\fa{b}) $3.6$. We show vortices
at two different scales: the grey objects are the smallest vortices identified
by positive isosurfaces of the second invariant $Q$ of the velocity gradient
tensor. To extract channel-half-width-scale vortices (i.e.~wall-attached
vortices in the outer layer), we apply the three-dimensional Gaussian filter
\citep{Motoori2019} with filter width $\ell = 0.2h$ to the fluctuation velocity.
We then evaluate the second invariant
$\mathrlap{\widetilde{\phantom{W}}}\,Q^{(\ell)}$ of the coarse-grained velocity
gradient tensor and show its positive isosurfaces in yellow. It is evident in
panel (\fa{a}) that the yellow outer-layer vortices are developed even in the
presence of particles with smaller $St_h$ ($=0.056$); whereas, in panel
(\fa{b}), these vortices are almost entirely attenuated due to particles with
larger $St_h$ ($=3.6$). 

\begin{figure}
	\centering
	\includegraphics[clip,width=0.75\textwidth]{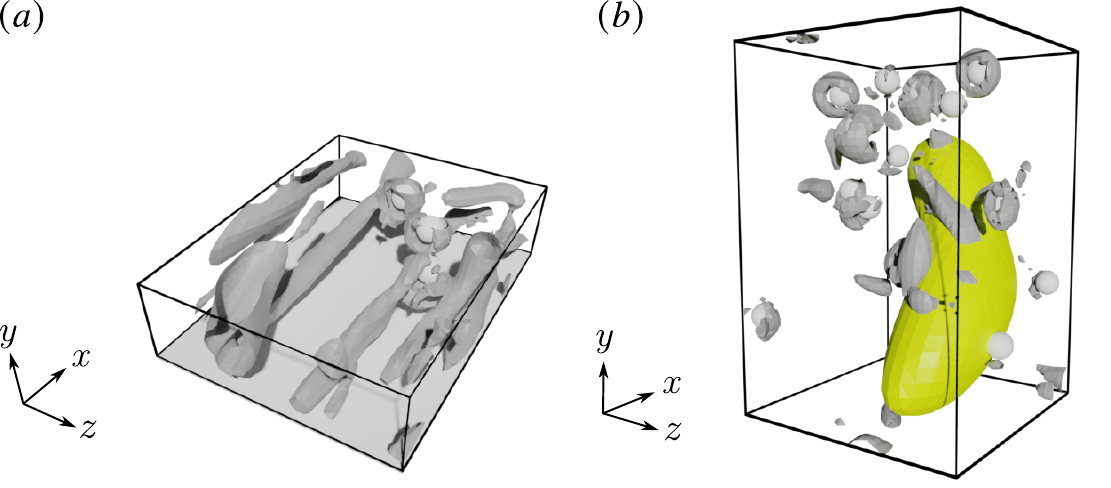}
	\caption{
	Magnification of subdomains in figure~\ref{fig:vis}(\fa{b}) for (\fa{a}) $0
	\leq y^+ \leq 50$ and (\fa{b}) $0.3 \leq y/h \leq 0.75$ (i.e.~$154 \leq y^+
	\leq 384$). Particles are depicted by white spheres.
    \label{fig:vis2}}
\end{figure}

It is also important to observe in figure~\ref{fig:vis}(\fa{b}) that vortex
rings are attached to particles. Figures~\ref{fig:vis2}(\fa{a}) and (\fa{b}) are
the magnifications of subdomains in figure~\ref{fig:vis}(\fa{b}) in the buffer
and outer layers, respectively. Here, particles are depicted by white spheres.
There are many vortex rings around particles in the both layers. Since $St_h$ of
these particles is larger than $1$, they can follow neither the outer-layer
vortices (with the longest time scale in the turbulence) nor the buffer-layer
vortices. The particles also cannot follow the mean flow because its time scale
is comparable to the turnover time of the wall-attached vortices at each height.
This explains the reason why these vortex rings are perpendicular to the
streamwise direction. If the particle Reynolds number were higher, the vortex
rings would be shed from the particles. In contrast, for $St_h \ll 1$ (see
figure~\ref{fig:vis}\fa{a}), there are no vortex rings in the outer layer. These
results therefore imply that the presence of particle-induced vortices results
in the reduction of turbulent vortices. As will be discussed in detail in the
following (see \S\:\ref{sec:mechanism}), these small vortices induced by
particles are indeed important because they produce the additional energy
dissipation, which is relevant to the turbulence attenuation.

\begin{figure}
	\centering
	\includegraphics[clip,width=\textwidth]{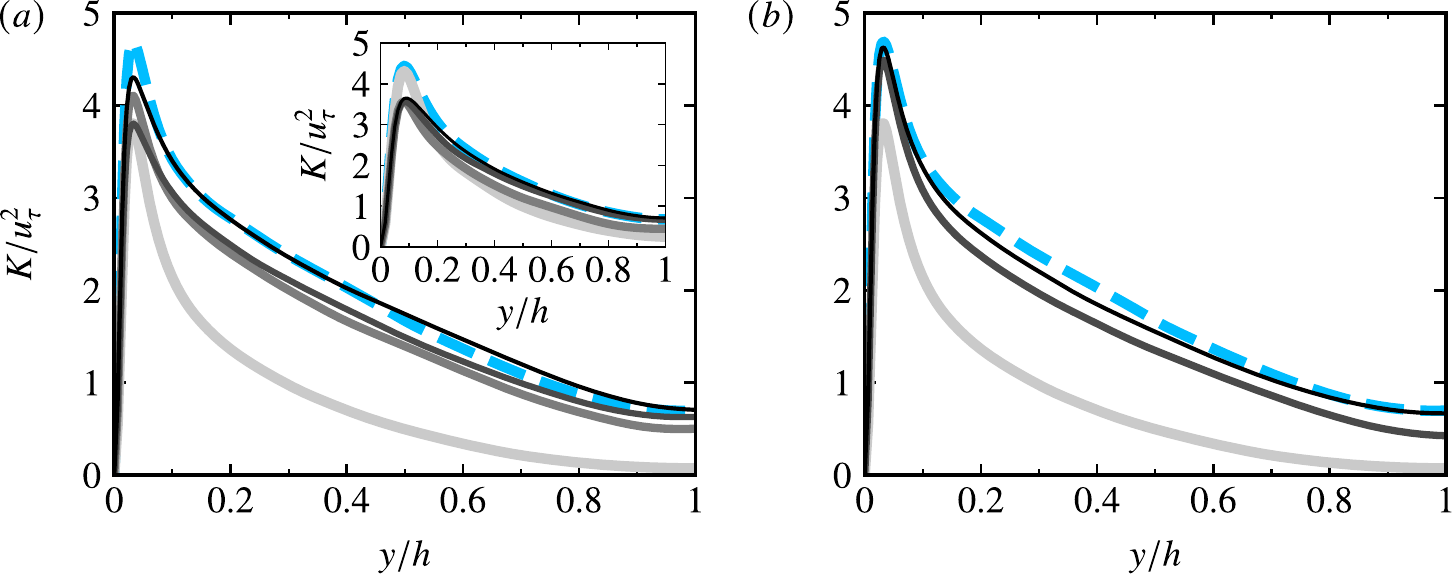}
	\caption{
	Wall-normal profiles of the mean turbulent kinetic energy $K(y)$ (\ref{eq:K})
	at $Re_\tau=512$. Panel (\fa{a}) shows the Stokes-number dependence
	[$St_h=0.056$ (black), $0.22$ (dark grey), $0.89$ (light grey) and $3.6$
	(very light grey)] for the common particle diameter $D/h=0.031$, while
	(\fa{b}) shows the particle-diameter dependence [$D/h=0.13$ (black), $0.063$
	(grey) and $0.031$ (light grey)] for the common Stokes number $St_h=3.6$.
	The blue dashed line shows the result for the single-phase flow. The inset
	in (\fa{a}) shows the results for $Re_\tau=180$ with $St_h=0.016$ (black),
	$0.63$ (dark grey), $2.5$ (light grey) and $10$ (very light grey).
    \label{fig:K-y}
	}
\end{figure}

\subsection{Turbulent kinetic energy\label{sec:energy}}

To quantify the degree of turbulence modulation at each height, we evaluate the
mean turbulent kinetic energy $K(y)$ defined as (\ref{eq:K}). As mentioned below
(\ref{eq:K}), when we take the spatial average, we only take into account the
fluid region outside particles. Figure~\ref{fig:K-y} shows the wall-normal
profiles of $K(y)$ for turbulence at $Re_\tau=512$. The blue dashed line
indicates the value $K_\times$ in the single-phase flow, where the subscript
$_\times$ denotes the value in the single-phase flow. First, let us look at
panel (\fa{a}), which shows the $St_h$-dependence for the smallest ($D/h=0.031$)
particles. Lighter and thicker lines indicate results for larger $St_h$. We see
in the panel that, irrespective of the height, turbulent kinetic energy is
attenuated more significantly for larger $St_h$. This is consistent with the
observation in figure~\ref{fig:vis} that particles with larger $St_h$ more
significantly attenuate outer-layer energetic vortices. More precisely, we can
observe in figure~\ref{fig:K-y}(\fa{a}) that as $St_h$ increases, $K$ is
attenuated from the lower height. For example, particles with $St_h=0.056$, that
is $St_+=28$ (the darkest and thinnest line), attenuate turbulence only for $y/h
\lesssim 0.1$ (i.e.~$y^+ \lesssim 50$). This is because the particle relaxation
time is too long for them to follow the swirls of the buffer-layer vortices
(whose time scale is in the order of $10\tau_+$) but short enough to follow the
larger wall-attached vortices such as the outer-layer vortices (because $St_h
\ll 1$). For the other $Re_\tau$, we have also confirmed that particles with
$St_+=28$ attenuate turbulent kinetic energy only in the buffer layer (figures
omitted). In contrast, particles with $St_h = 3.6$ (the lightest line) attenuate
the turbulent kinetic energy at any height. We observe similar trends with
respect to $St_h$ for $Re_\tau=180$ (see the inset in
figure~\ref{fig:K-y}\fa{a}) and for the results of smaller particles ($D^+
\approx 10$) in \citet{Shen2024}.

\begin{figure}
	\centering
	\includegraphics[clip,width=95mm]{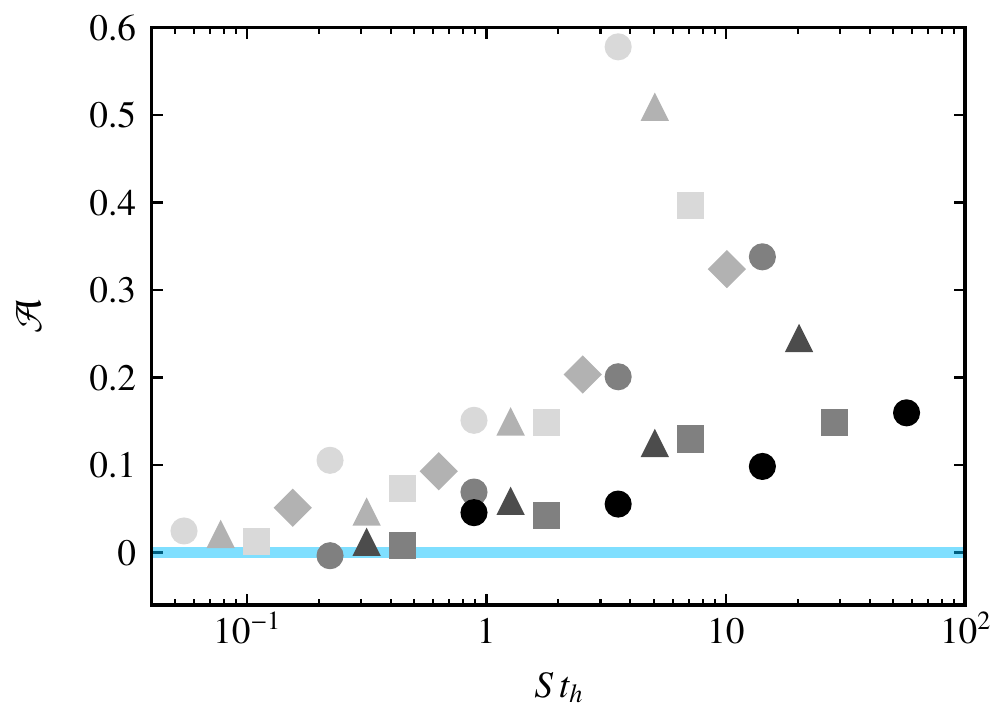}
	\caption{
	Average attenuation rate $\mathcal{A}$ (\ref{eq:ArK}) of turbulent kinetic
	energy as a function of $St_h$. Different symbols show the results for
	$Re_\tau = 512$ (circles), $360$ (triangles), $256$ (squares) and $180$
	(diamonds). The colour indicates $D/h$ (lighter symbols indicate smaller
	particles). 
    \label{fig:ArK}
	}
\end{figure}

Figure~\ref{fig:K-y}(\fa{b}) shows results for common $St_h=3.6$ but different
values of $D/h$ (darker and thinner lines indicate larger particles). Since
$St_h$ of these particles is larger than $1$, they attenuate turbulence at all
heights. It is however important to see in this panel that the attenuation rate
depends on the particle size; more precisely, it is larger for smaller diameters
for the fixed volume fraction of particles.

Thus, when particles cannot follow the wall-attached vortices (and therefore the
mean flow) at a given height $y$, the turbulent kinetic energy around $y$ is
attenuated. The degree of the turbulence attenuation is larger for larger Stokes
numbers (figure~\ref{fig:K-y}\fa{a}) and smaller diameters
(figure~\ref{fig:K-y}\fa{b}). 

Next, we consider spatially averaged quantities. Figure~\ref{fig:ArK} shows the
average attenuation rate, 
\begin{align}
\mathcal{A}
=
1-\frac{\langle K \rangle}{\langle K_\times \rangle} ,
\label{eq:ArK} 
\end{align}
of turbulent kinetic energy as a function of $St_h$. Here, $\langle \:\cdot\:
\rangle$ denotes the spatial average over the fluid region outside the
particles. Lighter colours indicate smaller $D/h$ and different symbols
represent different values of $Re_\tau$. We can see that irrespective of
$Re_\tau$, the average attenuation rate gets larger for larger $St_h$ and
smaller $D/h$. 

To explain these behaviours, we investigate how the kinetic energy transfers
from the mean flow to turbulence. Turbulent kinetic energy $K(y)$ at a given
height $y$ increases due to the production by the mean flow, transport from
other heights and work done by particles; on the other hand, it decreases due to
the transport to other heights and dissipation into heat. Since the average of
the spatial transport over the entire channel vanishes, the average budget of
turbulent kinetic energy $\langle K \rangle$ reads 
\begin{align}
\langle \mathcal{P}^{t \leftarrow M} \rangle
+
\langle \mathcal{W}^{t \leftarrow p} \rangle
=
\langle \epsilon \rangle 
, 
\label{eq:P+W=eps}
\end{align}
where $\mathcal{P}^{t \leftarrow M}$ is the turbulent energy production rate 
\begin{align}
\mathcal{P}^{t \leftarrow M}(y)
=
-\overline{u^\prime v^\prime} \dfrac{\d U}{\d y} 
\label{eq:P}
\end{align}
due to the mean flow and $\epsilon$ is the energy dissipation rate of
turbulence. In (\ref{eq:P+W=eps}), $\mathcal{W}^{t \leftarrow p}$
($=\mathcal{W}^{f \leftarrow p} - \mathcal{W}^{M \leftarrow p}$) is the rate of
work done by particles on turbulence. Here, $\mathcal{W}^{f \leftarrow p}$ is
the total rate of work by the force $\vb*{f}^{f \leftarrow p}$ that particles
exert on fluid: namely, $\mathcal{W}^{f \leftarrow p} = \overline{\vb*{f}^{f
\leftarrow p} \cdot \vb*{u}}$. By decomposing the velocity $\vb*{u}$ on the
particle surfaces into mean velocity $\vb*{U}$ in the fluid phase and its
fluctuation $\vb*{u}^\prime$, the total work can be separated into two
contributions: 
\begin{align}
\mathcal{W}^{f \leftarrow p} 
&=
\overline{\vb*{f}^{f \leftarrow p}}\cdot\vb*{U}
+
\overline{{{\vb*{f}}^{f \leftarrow p}}^\prime\cdot\vb*{u}^\prime}
\notag \\
&=
\mathcal{W}^{M \leftarrow p}
+
\mathcal{W}^{t \leftarrow p} .
\end{align}
In a statistically steady state, since the average kinetic energy of particles
remains constant, the average rate of work between fluid and particles vanishes,
i.e.~$\langle \mathcal{W}^{f \leftarrow p} \rangle = 0$. Therefore, when
$\langle \mathcal{W}^{t \leftarrow p} \rangle$ is positive, $\langle
\mathcal{W}^{M \leftarrow p} \rangle$ must be negative, meaning that particles
transfer the kinetic energy from the mean flow to turbulence. Thus, the energy
balance equation (\ref{eq:P+W=eps}) describes the spatially averaged energy
budget in turbulence with mean flow. In the following analyses, we investigate
how the energy budget is modulated due to particles on the basis of
(\ref{eq:P+W=eps}). Incidentally, we need a special care to evaluate the energy
dissipation rate $\epsilon$ in (\ref{eq:P+W=eps}) accurately in the DNS with IBM
(see Appendix~\ref{sec:eps}).

\subsection{Turbulent energy production by the mean flow\label{sec:production}}

\begin{figure}
	\centering
	\includegraphics[width=1.0\textwidth]{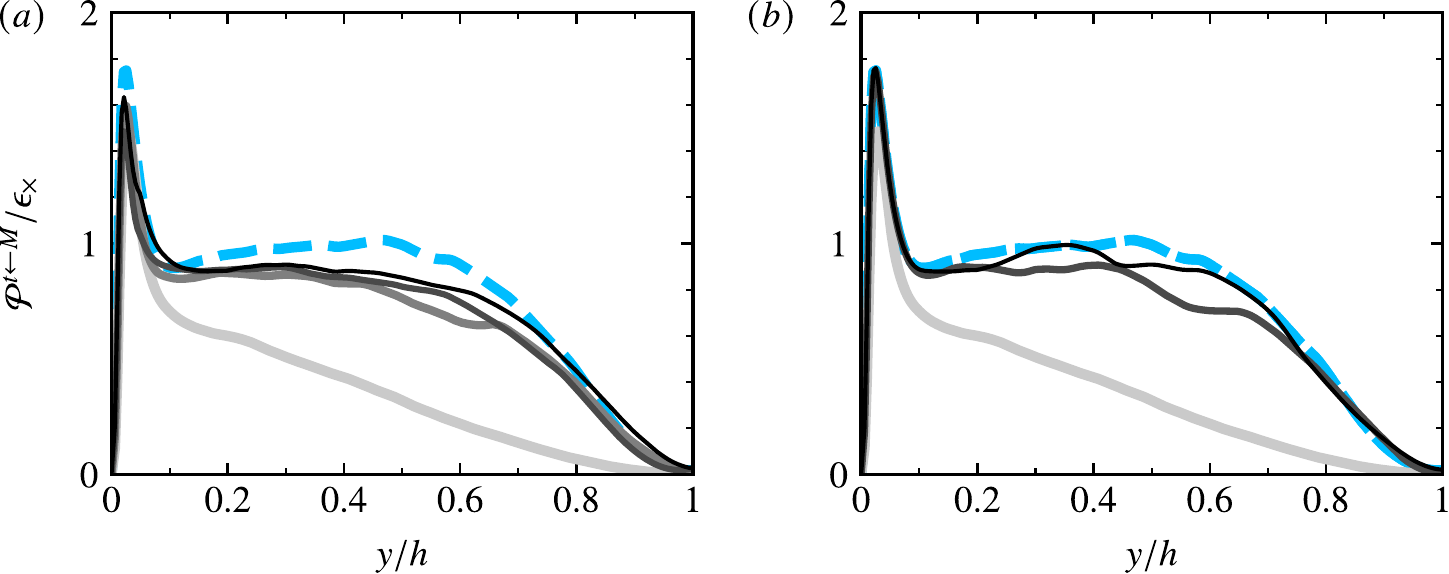}
	\caption{
	Wall-normal profiles of the mean turbulent energy production rate
	$\mathcal{P}^{t \leftarrow M}(y)$ (\ref{eq:P}) due to the mean flow at
	$Re_\tau=512$. The values are normalised by the mean turbulent energy
	dissipation rate $\epsilon_\times$ at each height in the single-phase flow.
	The lines in both panels indicate the same parameters as shown in
	figure~\ref{fig:K-y}. 
    \label{fig:P-y}}
\end{figure}

We first evaluate the turbulent energy production rate (\ref{eq:P}) due to the
mean flow. We show in figure~\ref{fig:P-y} the wall-normal profiles of
$\mathcal{P}^{t \leftarrow M}$ in turbulence at $Re_\tau=512$. Here, we
normalise $\mathcal{P}^{t \leftarrow M}(y)$ by the mean turbulent energy
dissipation rate $\epsilon_\times(y)$ of the single-phase turbulence. Particle
parameters in figure~\ref{fig:P-y} are the same as figure~\ref{fig:K-y}; namely,
panel (\fa{a}) shows results for the common $D/h$ ($=0.031$) but different
values of $St_h$, while (\fa{b}) shows those for the common $St_h$ ($=3.6$) but
different values of $D/h$. Figure~\ref{fig:P-y} shows that (i) when the Stokes
number is larger or (ii) the particle size is smaller under the condition that
the volume fraction is constant, the energy production rate tends to be more
reduced. It is also worth showing that the Reynolds stress ($-\overline{u^\prime
v^\prime}$) is also reduced (see Appendix~\ref{sec:U}). Since the Reynolds
stress is produced by wall-attached vortices \citep{Lozano-Duran2012,
Motoori2021}, turbulent kinetic energy, which is related to these energetic
vortices, is also attenuated.

\begin{figure}
	\centering
	\includegraphics[width=1.0\textwidth]{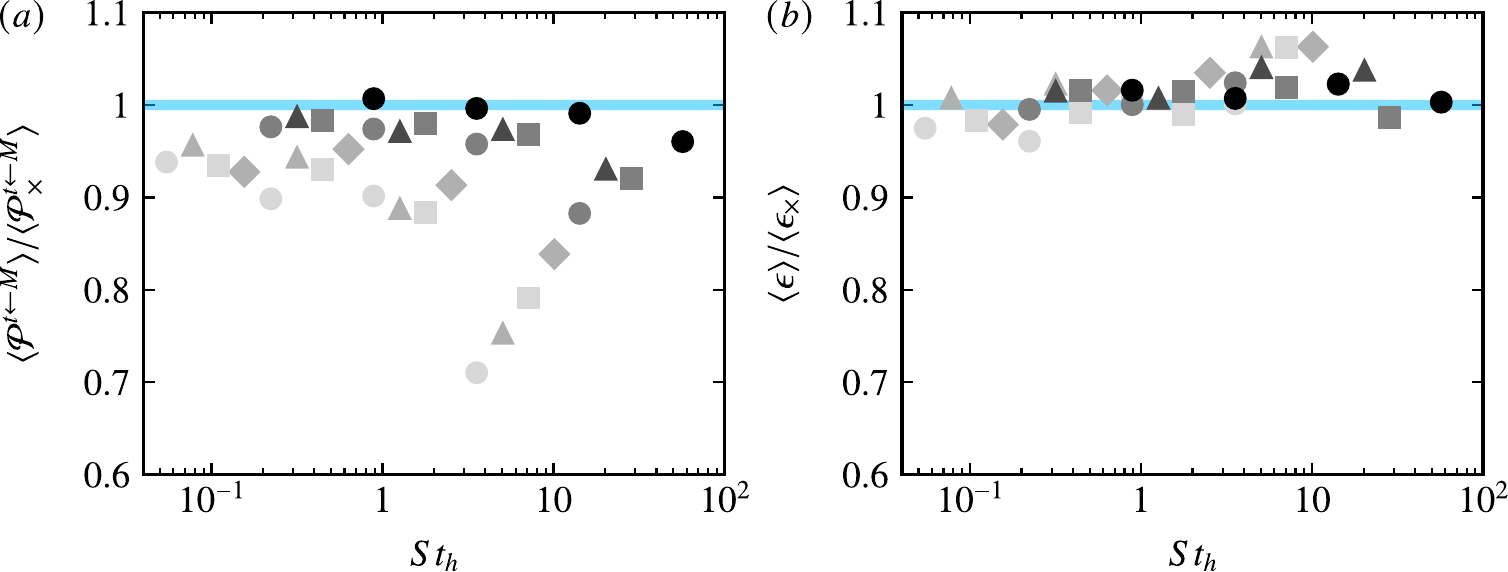}
	\caption{
	Spatial average of (\fa{a}) turbulent energy production rate $\langle
	\mathcal{P}^{t \leftarrow M} \rangle$ by the mean flow and (\fa{b}) turbulent energy
	dissipation rate $\langle \epsilon \rangle$ as functions of $St_h$. The
	values are normalised by those in the single-phase flow. The symbols are the
	same as in figure~\ref{fig:ArK}. 
    \label{fig:P_eps}
	}
\end{figure}

Next, we show in figure~\ref{fig:P_eps}(\fa{a}) the spatial average $\langle
\mathcal{P}^{t \leftarrow M} \rangle$ of the energy production rate normalised
by the value $\langle \mathcal{P}^{t \leftarrow M}_\times \rangle$ in the
single-phase flow as a function of $St_h$. The symbols are the same as in
figure~\ref{fig:ArK}; namely, those in lighter colours indicate smaller $D/h$,
and different shapes represent different values of $Re_\tau$. We see that the
average production rate gets smaller for (i) larger $St_h$ and (ii) smaller
$D/h$. These behaviours are similar to those for turbulent kinetic energy shown
in figure~\ref{fig:ArK}.

It is also important to observe that the turbulent energy dissipation rate
$\langle \epsilon \rangle$ is not modulated by particles as much as $\langle
\mathcal{P}^{t \leftarrow M} \rangle$. Figure~\ref{fig:P_eps}(\fa{b}) shows
$\langle \epsilon \rangle$ normalised by $\langle \epsilon_\times \rangle$ as a
function of $St_h$. Note that $\epsilon$ includes the two contributions from the
energy dissipation through the energy cascade and from the wake behind added
particles. We see that $\langle \epsilon \rangle / \langle \epsilon_\times
\rangle$ is close to unity. This result means that the average energy input rate
$\langle I^M \rangle$ minus the average energy dissipation rate $\langle
\epsilon^M \rangle$ of the mean flow is not significantly altered from that of
the single-phase flow. More concretely, $\langle \mathcal{I}^M \rangle$ and
$\langle \epsilon^M \rangle$ show similar monotonic trends with respect to
$St_h$; see Appendix~\ref{sec:MKE}.

\subsection{Additional energy dissipation due to particles\label{sec:mechanism}}

We have demonstrated in figures~\ref{fig:P-y} and \ref{fig:P_eps}(\fa{a}) that
particles can reduce the energy production rate $\mathcal{P}^{t \leftarrow M}$,
which relates to the attenuation of turbulent kinetic energy. We have also shown
in figures~\ref{fig:vis} and \ref{fig:vis2} that particle-induced vortices seem
relevant to turbulence attenuation. In this subsection, we discuss how these
vortices contribute to the reduction of $\mathcal{P}^{t \leftarrow M}$.

\begin{figure}
	\centering
	\includegraphics[width=1.0\textwidth]{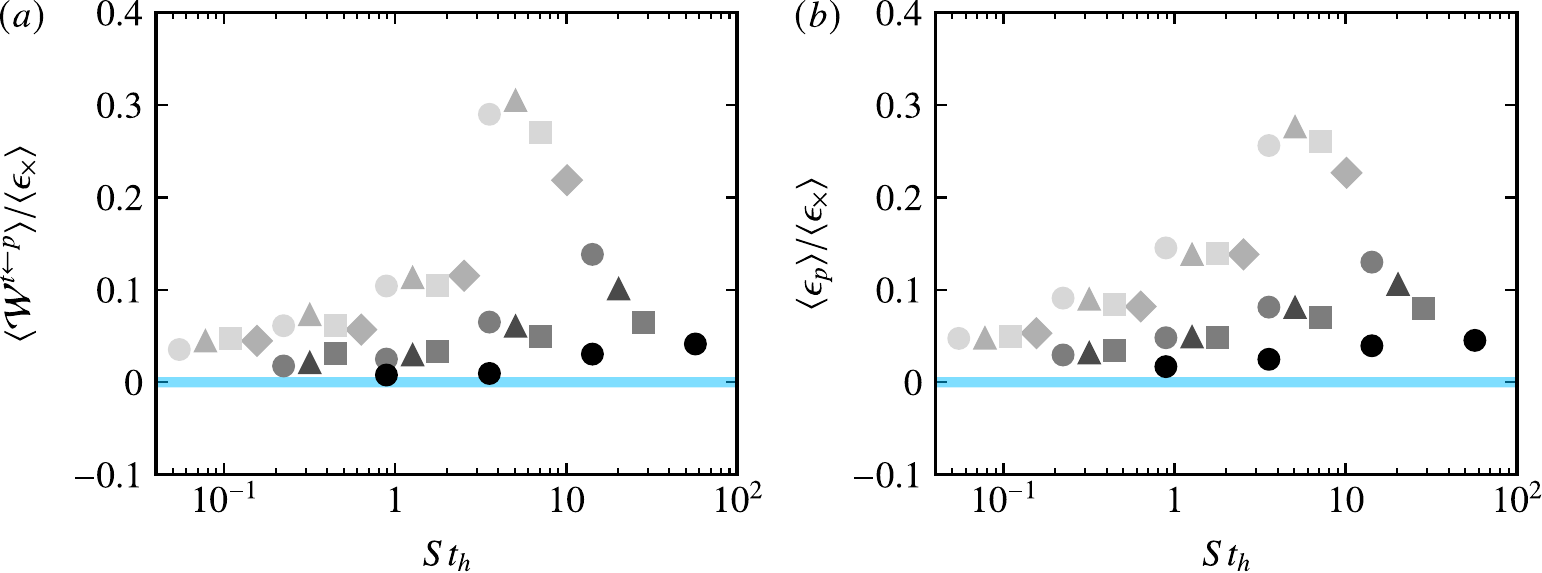}
	\caption{
	Spatial average of (\fa{a}) rate of work $\mathcal{W}^{t \leftarrow p}$ done
	by particles on turbulence evaluated by (\ref{eq:P+W=eps}) and (\fa{b})
	energy dissipation rate $\epsilon_p$ around particles defined with
	(\ref{eq:epsp}) as functions of $St_h$. The values are normalised by the
	spatial average $\langle \epsilon_\times \rangle$ of turbulent energy
	dissipation rate in the single-phase flow. The symbols are the same as in
	figure~\ref{fig:ArK}. 
    \label{fig:epsp}
	}
\end{figure}

For the spatial average in the single-phase flow, the energy production rate due
to the mean flow is balanced by the energy dissipation rate, i.e.~$\langle
\mathcal{P}^{t \leftarrow M} \rangle = \langle \epsilon \rangle$. However,
particles can break this balance due to the rate of work $\mathcal{W}^{t
\leftarrow p}$ done by particles on turbulence, as described in
(\ref{eq:P+W=eps}). Figure~\ref{fig:epsp}(\fa{a}) shows the average of
$\mathcal{W}^{t \leftarrow p}$ evaluated by (\ref{eq:P+W=eps}). We see that
$\langle \mathcal{W}^{t \leftarrow p} \rangle$ gets larger for particles with
larger $St_h$ and smaller $D/h$. This implies that particles play a role in
transferring turbulent kinetic energy from the mean flow to turbulence. As a
result, vortex rings shown in figures~\ref{fig:vis} and \ref{fig:vis2} are
generated. Then, the turbulent kinetic energy associated with these vortices is
rapidly dissipated, leading to a substantial energy dissipation rate
$\epsilon_p$ around the particles. Therefore, combining the local balance
$\mathcal{W}^{t \leftarrow p} = \epsilon_p$ with (\ref{eq:P+W=eps}), we can
describe the energy production rate as
\begin{align}
\langle \mathcal{P}^{t \leftarrow M} \rangle
=
\langle \epsilon \rangle
-
\langle \epsilon_p \rangle 
\label{eq:P_balance} .
\end{align}
Since $\langle \epsilon \rangle =
\langle \epsilon_\times \rangle$ approximately holds in the present system as
shown in figure~\ref{fig:P_eps}(\fa{b}), we can rewrite (\ref{eq:P_balance}) as
\begin{align}
1
-
\frac{
\langle \mathcal{P}^{t \leftarrow M} \rangle}{\langle \mathcal{P}^{t \leftarrow M}_\times \rangle}
=
\frac{
\langle \epsilon_p \rangle }{ \langle \epsilon_\times \rangle}.
\label{eq:ArP}
\end{align}
This equation means that the degree of the reduction of $\langle \mathcal{P}^{t
\leftarrow M} \rangle$ is proportional to the energy dissipation rate $\langle
\epsilon_p \rangle$ due to vortices induced by particles. 

\begin{figure}
	\centering
	\includegraphics[width=1.0\textwidth]{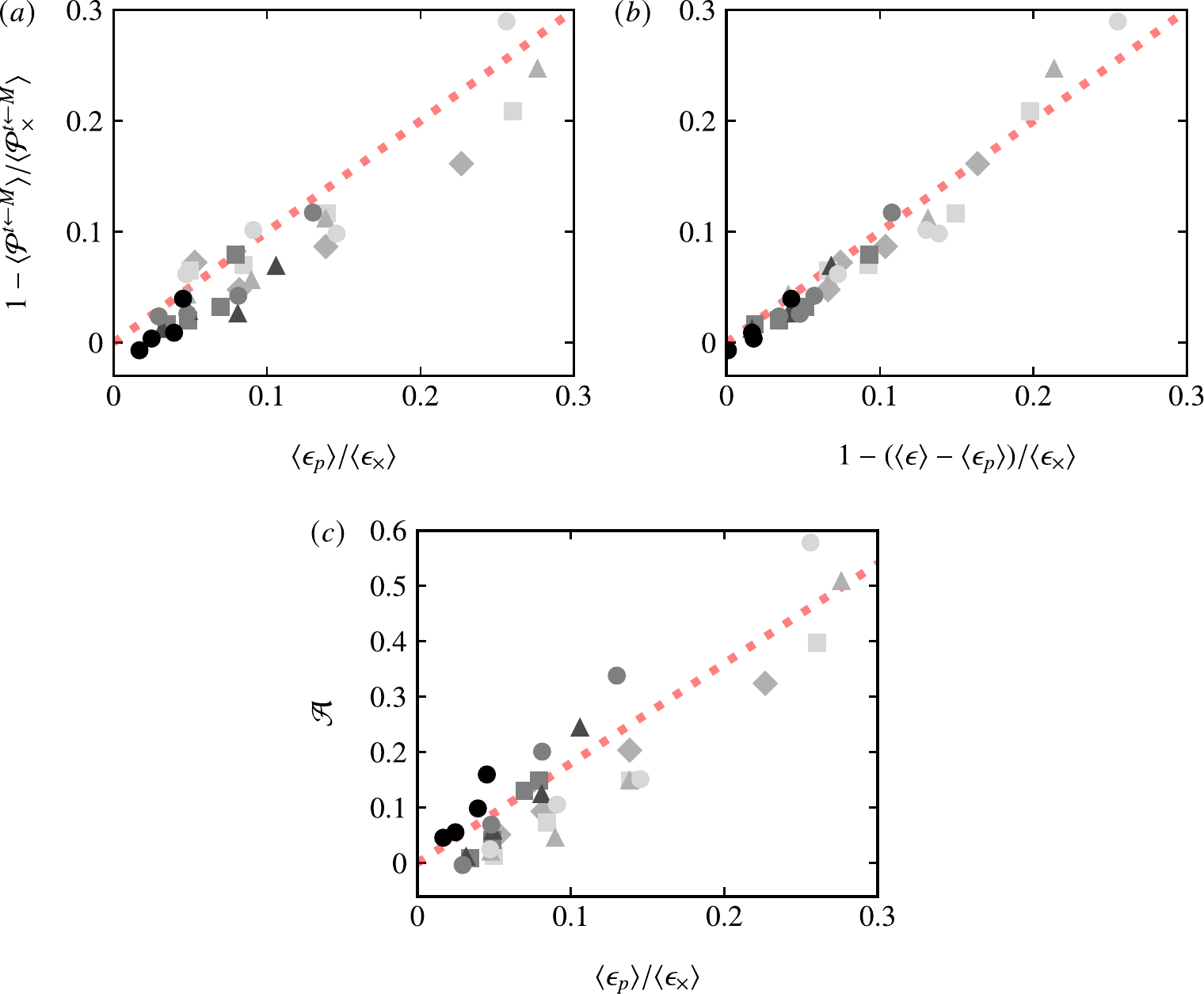}
	\caption{
	Average attenuation rate of turbulent energy production rate $\langle
	\mathcal{P}^{t \leftarrow M} \rangle$ by the mean flow as functions of
	(\fa{a}) $\langle \epsilon_p \rangle/\langle \epsilon_\times \rangle$ and
	(\fa{b}) $1-(\langle \epsilon \rangle-\langle \epsilon_p \rangle)/\langle
	\epsilon_\times \rangle$. (\fa{c}) Average attenuation rate $\mathcal{A}$ of
	turbulent kinetic energy as a function of $\langle \epsilon_p
	\rangle/\langle \epsilon_\times \rangle$. The symbols are the same as in
	figure~\ref{fig:ArK}. The proportional coefficients of the red dotted lines
	are (\fa{a}, \fa{b}) $1$ and (\fa{c}) $1.8$.
    \label{fig:Ar-epsp}
	}
\end{figure}

To verify the above argument, we show in figure~\ref{fig:epsp}(\fa{b}) the
spatial average $\langle \epsilon_p \rangle$ of the energy dissipation rate
$\epsilon_p$ around particles normalised by $\langle \epsilon_\times \rangle$.
Here, to estimate $\langle \epsilon_p \rangle$, we numerically compute the local
average of the turbulent energy dissipation rate around all particles: 
\begin{align}
\langle \epsilon_p \rangle =
\frac{1}{V_{box}}\lk
\Bigl\langle \int_{\Omega_\circledcirc} \epsilon(\vb*{x}^\prime,t) \: \mathrm{d}
V^\prime \Bigl\rangle -
V_\circledcirc \: \langle \epsilon \rangle
\rk .
\label{eq:epsp}
\end{align}
Here, $\Omega_\circledcirc$ denotes the region between a spherical particle and
another sphere with diameter $3D$ concentric with the particle \citep{Sun2024},
$V_{\circledcirc}$ is the volume of $\Omega_\circledcirc$, and $V_{box}$ is the
volume of the computational box. The first term on the right-hand side of
(\ref{eq:epsp}) captures the local average of the energy dissipation rate
$\epsilon$ around particles, where we evaluate $\epsilon$ by (\ref{eq:eps}) in
Appendix~\ref{sec:eps}. Since this quantity includes the dissipation rates due
to not only the vortices created around particles but also vortices generated by
energy cascade or the mean shear, we subtract the latter contribution expressed
by the second term. Figure~\ref{fig:epsp}(\fa{b}) shows that $\langle \epsilon_p
\rangle$ gets larger for larger $St_h$ or smaller $D/h$, which is similar to the
behaviour of $\langle \mathcal{W}^{t \leftarrow p} \rangle$ in (\fa{a}). We can
explain these dependences by noting that the local energy dissipation rate per
unit mass around a single particle is approximated by $|\Delta u|^3/D$, where
$\Delta u$ is the magnitude of the relative velocity. Hence, when (i) the Stokes
number is larger, and therefore the relative velocity is larger or (ii) the
diameter is smaller, $\epsilon_p$ gets larger under the condition the particle
volume fraction is fixed. Recall that figure~\ref{fig:P_eps}(\fa{a}) shows a
similar tendency for the average reduction rate $\langle \mathcal{P}^{t
\leftarrow M} \rangle$. To verify this similarity, we show $1-\langle
\mathcal{P}^{t \leftarrow M} \rangle/\langle \mathcal{P}^{t \leftarrow M}_\times
\rangle$ as a function of $\langle \epsilon_p \rangle / \langle \epsilon_\times
\rangle$ in figure~\ref{fig:Ar-epsp}(\fa{a}). The data approximately collapse on
the red dotted line. This implies that we can describe the average reduction
rate of the energy production $\langle \mathcal{P}^{t \leftarrow M} \rangle$ due
to the mean flow in terms of the additional energy dissipation rate $\langle
\epsilon_p \rangle$ by particles, as indicated by (\ref{eq:ArP}). Incidentally,
although some plots lie beneath the red line, this deviation arises from the
slight modulation of $\langle \epsilon \rangle$ from the single-phase flow (see
figure~\ref{fig:P_eps}\fa{b}). Taking this effect into account, when we replot
the attenuation rate of $\langle \mathcal{P}^{t \leftarrow M} \rangle$ as a
function of the energy flux $\langle \epsilon \rangle - \langle \epsilon_p
\rangle$ in figure~\ref{fig:Ar-epsp}(\fa{b}), the plots collapse better.

Moreover, the additional energy dissipation rate is also important in describing
the average attenuation rate of turbulent kinetic energy. We show in
figure~\ref{fig:Ar-epsp}(\fa{c}) the average attenuation rate $\mathcal{A}$,
defined as (\ref{eq:ArK}), of turbulent kinetic energy. Although the turbulence
attenuation rate depends on $St_h$, $D/h$ and $Re_\tau$ (see
figure~\ref{fig:ArK}), when plotting $\mathcal{A}$ as a function of $\langle
\epsilon_p \rangle$/$\langle \epsilon_\times \rangle$, the data approximately
collapse on the red dotted line. We conclude therefore that the turbulence
attenuation rate is related to the additional energy dissipation rate $\langle
\epsilon_p \rangle$ due to particles. This is consistent with previous studies
on modulation of wall-bounded turbulence by pointwise particles \citep{Zhao2013}
and fixed particles with restricted their motion \citep{Peng2019b}, which
emphasised the importance of the additional energy dissipation around particles.
Incidentally, in figure~\ref{fig:Ar-epsp}(\fa{c}), we observe the
proportionality between $\mathcal{A}$ and $\langle \epsilon_p \rangle / \langle
\epsilon_\times \rangle$, indicating that $\langle \mathcal{P}^{t \leftarrow M}
\rangle$ is proportional to $\langle K \rangle$. However, there may be a better
expression for the relationship between $\mathcal{P}^{t \leftarrow M}$ and $K$
for higher-Reynolds-numbers turbulence. We further discuss this in
\S\:\ref{sec:Ar}.

We summarise the mechanism of attenuation of wall-bounded turbulence. When
particles cannot follow the fluid motions and create the small vortices around
them (figures~\ref{fig:vis} and \ref{fig:vis2}), turbulent kinetic energy $K$ is
attenuated (figure~\ref{fig:K-y}). This occurs because these particle-induced
vortices acquire a part of the kinetic energy from the mean flow
(figure~\ref{fig:epsp}\fa{a}), and produce the additional energy dissipation
rate (figure~\ref{fig:epsp}\fa{b}), thereby preventing energy transfer rate
$\mathcal{P}^{t \leftarrow M}$ from the mean flow to turbulent vortices
(figures~\ref{fig:P-y} and \ref{fig:P_eps}\fa{a}). This is the reason why we can
describe the average attenuation rates of $\mathcal{P}^{t \leftarrow M}$ and $K$
in terms of the additional energy dissipation rate $\langle \epsilon_p \rangle$
due to particles (figure~\ref{fig:Ar-epsp}).

Incidentally, it is also worth mentioning that the Reynolds stress is also
reduced (figure~\ref{fig:U}\fa{b} in Appendix~\ref{sec:U}), leading to a
reduction of the frictional drag coefficient. This is because the mean bulk
velocity increases (figure~\ref{fig:U}\fa{a}) as a consequence of the reduction
of fluid momentum transfer in the wall-normal direction. We show this result in
Appendix~\ref{sec:Cf}.

\section{Prediction of turbulence attenuation rate\label{sec:discussion}}

In this section, we propose a method to predict the degree of turbulence
attenuation in terms of the given property of particles. To this end, we must
predict the additional energy dissipation rate $\langle \epsilon_p \rangle$,
which determines the turbulence attenuation rate, without the direct evaluation
(\ref{eq:epsp}) as in the previous section.

\subsection{Estimation of particle energy dissipation\label{sec:epsp-outer}}

We first consider the magnitude of the force exerted on a particle by the fluid:
\begin{align}
F^{p \leftarrow f} = \frac{1}{2} C_D \rho_f |\Delta u|^2 A,
\end{align}
where $C_D$ is the drag coefficient, $A$ ($=\pi D^2/4$) is the cross-sectional
area of the sphere, and $\Delta u$ is the magnitude of the relative velocity.
The particle subjected to this force leads to the energy dissipation at the rate
of $F^{p \leftarrow {f}} \Delta u$. Therefore, the energy dissipation rate due to
all the particles in the system can be expressed as
\begin{align}
\langle \epsilon_p^\# \rangle = \frac{3}{4} \Lambda_0 C_D \frac{|\Delta u|^3}{D}.
\label{eq:epsp_sh}
\end{align}
Thus, to obtain $\langle \epsilon_p^\# \rangle$, we need to estimate the
relative velocity $\Delta u$. We discuss the estimation of $\Delta u$ in the
next subsection.

Moreover, by estimating the average energy dissipation rate in the single-phase
flow as $\langle \epsilon_\times^\# \rangle = u_L^3/h$ with $u_L$($\sim
\hspace{-1mm}u_\tau$) being the characteristic velocity at the largest scale in
the outer layer, we obtain
\begin{align}
\frac{\langle \epsilon_p^\# \rangle}{\langle \epsilon_\times^\#
\rangle}
=
\frac{3}{4} \Lambda_0 C_D \frac{h}{D} \frac{{|\Delta u|^3}}{u_L^3} .
\label{epsp_sh_ratio}
\end{align}
For the drag coefficient $C_D$, we use the experimental law
\citep{Schiller1933}: 
\begin{align}
C_D  = \frac{24}{Re_D}(1+0.15Re_D^{0.687}) , 
\label{eq:C_D}
\end{align}
where $Re_D$ ($=|\Delta u| D/\nu$) is the particle Reynolds number. Thus, we can
describe the additional energy dissipation rate (\ref{epsp_sh_ratio}) by using
the particle parameters ($\Lambda_0$ and $D/h$) and the relative velocity
$\Delta u$. 

\subsection{Relative velocity\label{sec:du-outer}}

\begin{figure}
	\centering
	\includegraphics[width=95mm]{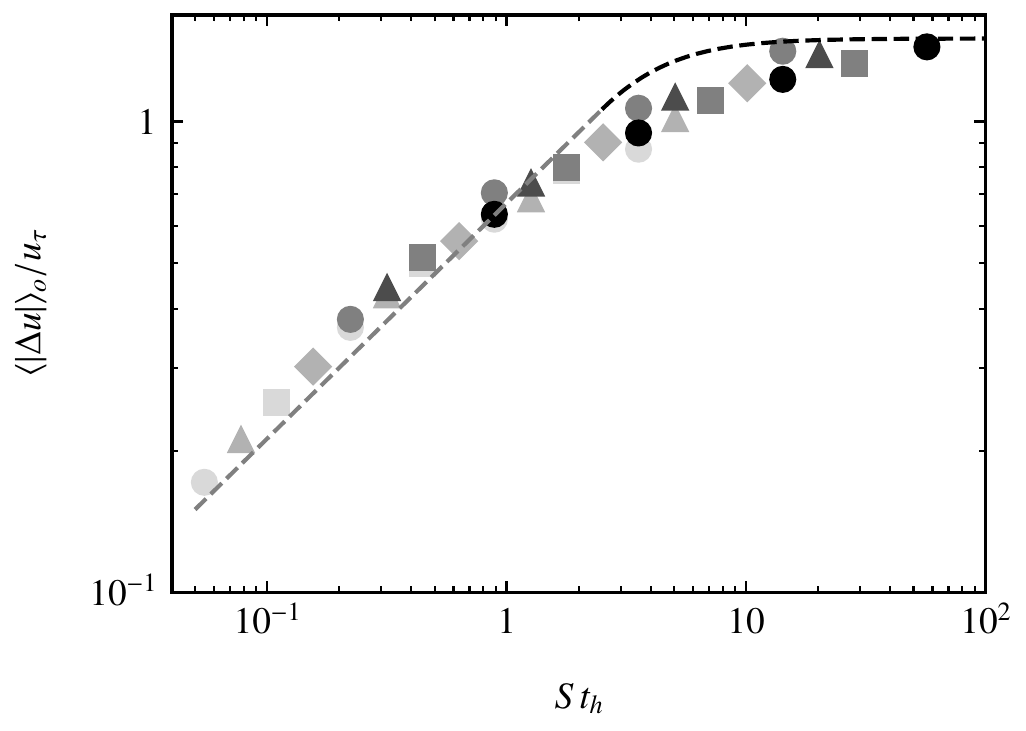}
	\caption{
	Relative velocity $\langle | \Delta u | \rangle_o$ between particles and their
	surrounding fluid averaged in the outer layer. The symbols are the same as
	in figure~\ref{fig:ArK}. The grey and black dashed lines are evaluation by
	(\ref{eq:bala-ii}) and (\ref{eq:bala-iii}), respectively, with coefficients
	of $u_L=1.5u_\tau$ and $\tau_L=2.5\tau_h$. 
    \label{fig:du}}
\end{figure}

We show in figure~\ref{fig:du} the relative velocity $\langle |\Delta {u}|
\rangle_{o}$ averaged in the outer layer ($y/h \geq 0.3$) as a function of
$St_h$. Note that since $\Delta u$ strongly depends on height $y$, we focus on
the outer layer where $\Delta u$ weakly depends on $y$ (see
Appendix~\ref{sec:du-y}). Here, we evaluate the relative velocity as $\Delta u =
\langle {u}\rangle_{\text{\textcircled{$p$}}} - u_{p}$, where we define the
surrounding fluid velocity $\langle {u}\rangle_{\text{\textcircled{$p$}}}$ for
each particle by the average fluid velocity on the surface of a sphere with
diameter $2D$ concentric with the particle \citep{Kidanemariam2013,
Uhlmann2017}. Symbols in lighter colours indicate smaller $D/h$, and different
shapes represent different values of $Re_\tau$. We see in the figure that $St_h$
determines the relative velocity irrespective of the other parameters. We also
see that the functional forms of the relative velocity are well approximated by
the dashed lines, which are defined as
\begin{subequations}
\label{eq:bala}
\begin{empheq}[left = {{|\Delta {u}|}
= \empheqlbrace \,}, right = {}]{align}
& \frac{(\tau_p/\tau_L)^{\frac12}}{\sqrt{2}} \: u_L
\hspace{-14mm}
&(\tau_\eta \leq \tau_p \leq \tau_L),\: \label{eq:bala-ii} \\
& \frac{\tau_p/\tau_L}{\sqrt{1+(\tau_p/\tau_L)^2}} \: u_L 
\hspace{-14mm}
&(\tau_p > \tau_L).\: \label{eq:bala-iii}
\end{empheq}
\end{subequations}
Here, (\ref{eq:bala-ii}) holds when the particle relaxation time $\tau_p$ is
longer than the Kolmogorov time $\tau_\eta$ and shorter than the integral time
$\tau_L$, whereas (\ref{eq:bala-iii}) holds for $\tau_p > \tau_L$. We can derive
(\ref{eq:bala}) by assuming pointwise heavy particles based on the argument by
\citet{Balachandar2009} as follows. As considered in our previous studies
\citep{Oka2021, Motoori2022}, we first assume that the motion of particles with
$\tau_p$ is independent of fluid motions smaller than $\ell$, where $\ell$ is
the length scale such that the turnover time $\tau^{(\ell)}$ ($=\langle
\epsilon_\times \rangle^{-\frac{1}{3}} \ell^{\frac{2}{3}}$) of vortices with
size $\ell$ is approximately $\tau_p$, i.e.~$\ell = \langle \epsilon_\times
\rangle^\frac12 \tau_p^\frac32$. Then, considering a particle in the oscillating
flow with the frequency $\tau^{(\ell)}$ \citep{Balachandar2009}, we obtain
\begin{align}
\label{eq:OWG}
|\Delta u|
&=
\frac{\tau_p/\tau^{(\ell)}}
{\sqrt{1+(\tau_p/\tau^{(\ell)})^2}}
\: 
\widetilde{u}^{({\ell})} .
\end{align}
Here, $\widetilde{u}^{(\ell)}$ is the fluid velocity coarse-grained at scale
$\ell$. For $\tau_\eta \leq \tau_p \leq \tau_L$, when assuming that the relative
velocity is determined by vortices whose turnover time is comparable to the
particle relaxation time (i.e.~$\tau^{(\ell)} = \tau_p$), we can obtain
(\ref{eq:bala-ii}). In this derivation, we use $\widetilde{u}^{(\ell)} =
(\tau^{(\ell)}/\tau_L)^{\frac12} u_L$. On the other hand, we can obtain
(\ref{eq:bala-iii}) for $\tau_p > \tau_L$ by assuming that the relative velocity
is determined by the largest-scale vortices (i.e.~$\tau^{(\ell)} = \tau_L$). 

Figure~\ref{fig:du} shows that (\ref{eq:bala}) is in good agreement with our DNS
data. Here, we choose the parameters $u_L$ and $\tau_L$ in (\ref{eq:bala}) as in
the order of $u_\tau$ and $\tau_h$, respectively, so that the curves expressed
by (\ref{eq:bala}) fits our data. Incidentally, the relative velocity for
$\tau_p < \tau_\eta$ is determined by vortices at the Kolmogorov time scale
(i.e.~$\tau^{(\ell)}= \tau_\eta$), although in the present study, we do not
simulate particles with $\tau_p < \tau_\eta$.

Incidentally, it is not always possible to estimate the relative velocity using
(\ref{eq:bala}). This evaluation is valid when the particle diameter is smaller
than the length scale $\ell$ of vortices with the turnover time $\tau^{(\ell)}
\approx \tau_p$, i.e.,
\begin{subequations}
\label{eq:condition}
\begin{empheq}[
left = { 
D \lesssim \ell
\quad \Leftrightarrow \quad 
\dfrac{D}{h} \lesssim
\empheqlbrace 
\, }, right = {}]{align}
& (\tau_p/\tau_L)^{\frac{3}{2}} \approx St_h^{\frac{3}{2}} \hspace{-14mm}
&(\tau_\eta \leq \tau_p \leq \tau_L),\: \\
& 1 \hspace{-14mm} &(\tau_p > \tau_L).\: 
\end{empheq} 
\end{subequations} 
The examined particles satisfy this condition.

\subsection{Prediction of turbulence attenuation rate\label{sec:Ar}}

\begin{figure}
	\centering
	\includegraphics[width=95mm]{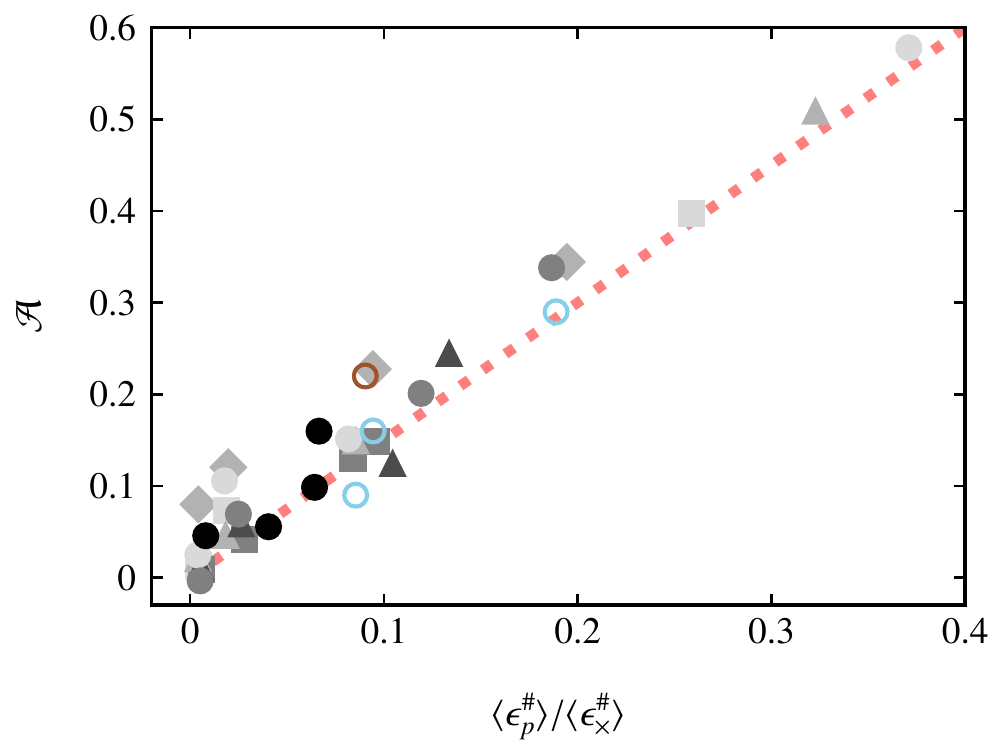}
	\caption{
	Average attenuation rate $\mathcal{A}$ of turbulent kinetic energy as a
	function of $\langle \epsilon_p^\# \rangle/\langle \epsilon_\times^\#
	\rangle$, which is estimated by (\ref{eq:epsp_pre}) using only particle
	parameters. Particle parameters for the closed symbols are the same as in
	figure~\ref{fig:ArK}. The open circles show the turbulence attenuation rate
	at the channel centre measured in experiments by \citet{Kulick1994}. The
	brown and light blue symbols indicate the results for coppers and glasses
	(with different volume fractions $\Lambda_0$), respectively. The
	proportional coefficient of the red dotted line is $1.5$.
    \label{fig:Ar-epsp_sh}}
\end{figure}

By substituting the relative velocity prediction (\ref{eq:bala})
into (\ref{epsp_sh_ratio}), we can estimate the additional energy
dissipation rate by
\begin{subequations}
\label{eq:epsp_pre}
\begin{empheq}[left = { 
\dfrac{\langle \epsilon_p^\# \rangle}{\langle \epsilon_\times^\# \rangle}
= \empheqlbrace \,}, right = {}]{align}
& 
\frac{3}{4} \Lambda_0 C_D \frac{h}{D} \frac{(\tau_p/\tau_L)^{\frac32}}{2\sqrt{2}} 
\hspace{-14mm}
&(\tau_\eta \leq \tau_p \leq \tau_L),\: \label{eq:epsp_pre-ii} \\
&
\frac{3}{4} \Lambda_0 C_D \frac{h}{D} \frac{(\tau_p/\tau_L)^{3}}{{(1+(\tau_p/\tau_L)^2})^\frac32}
\hspace{-14mm}
&(\tau_p > \tau_L).\: \label{eq:epsp_pre-iii}
\end{empheq}
\end{subequations}
We plot in figure~\ref{fig:Ar-epsp_sh} the average attenuation rate
$\mathcal{A}$ as a function of $\langle \epsilon_p^\# \rangle/\langle
\epsilon_\times^\# \rangle$. We see that our DNS data collapse onto a single
line, which is approximated by the red dotted line. In this figure, we also plot
the experimental results \citep{Kulick1994} for turbulent channel flow. The
brown and light blue open symbols indicate the attenuation rates at the channel
centre by copper and glass particles, respectively. Despite the different
particle types and volume fractions, the experimental data align closely with
our DNS results. Thus, we can use (\ref{eq:epsp_pre}) to estimate the additional
energy dissipation rate $\langle \epsilon_p^\# \rangle/\langle
\epsilon_\times^\# \rangle$, and by using this estimation, we can describe the
turbulence attenuation rate.

Before concluding this article, we discuss the relevance to the study on
periodic turbulence by \citet{Oka2022}. They derived the formula 
\begin{align}
1-\left( 1-\frac{\mathcal{A}}{1+\alpha} \right)^{\frac{3}{2}} 
= 
\frac{\langle \epsilon_p \rangle}{\langle \epsilon_\times \rangle} 
\label{eq:OG22}
\end{align}
for describing the turbulence attenuation rate, and then verified it using their
DNS results of periodic turbulence. Here, $\alpha$ is the ratio of the kinetic
energy of the mean flow to turbulent energy for the single-phase flow. When
deriving this formula, they first assumed that the additional energy dissipation
rate bypasses the energy cascade. Then, they used \citet{Taylor1935}'s
dissipation law to relate average turbulent kinetic energy $\langle K \rangle$
to its dissipation rate $\langle \epsilon \rangle$. We may use a similar
relation between $K$ and $\epsilon$ ($= \mathcal{P}^{t \leftarrow M}$) in the
log layer for wall turbulence at sufficiently high Reynolds numbers. However,
since the present turbulence does not have a large scale separation to discuss
the buffer, log and outer layers individually, we have argued the spatially
averaged quantities. Nevertheless, our DNS results show that the turbulence
attenuation rate increases monotonically with respect to $\langle \epsilon_p
\rangle / \langle \epsilon_\times \rangle$ (see figure~\ref{fig:Ar-epsp}\fa{c}).
Moreover, we have demonstrated in figure~\ref{fig:Ar-epsp_sh} that $\langle
\epsilon_p \rangle / \langle \epsilon_\times \rangle$ can be estimated based
solely on particle parameters using (\ref{eq:epsp_pre}). This estimation can be
applicable to developed turbulence laden with small particles satisfying
(\ref{eq:condition}) in a dilute regime. 

\section{Conclusions}

To investigate the attenuation of wall-bounded turbulence due to heavy small
particles in a dilute regime, we have conducted DNS of turbulent channel flow
laden with finite-size solid particles. Fixing the small volume fraction
($0.82\,\%$) of particles, we change the particle diameter, particle relaxation
time and turbulence Reynolds number (table~\ref{table:parameter}). 
The conclusions of the present study are as follows.

When particles cannot follow the ambient fluid; namely, when the particle
relaxation time is longer than the swirling time of the wall-attached vortices
at the particles' existing height, vortex rings are created by the particles
(figures~\ref{fig:vis} and \ref{fig:vis2}). The presence of such
particle-induced vortices results in a significant turbulence attenuation
(figures~\ref{fig:K-y} and \ref{fig:ArK}). This is because they produce the
additional energy dissipation (figure~\ref{fig:epsp}\fa{b}), which bypasses the
energy production from the mean flow to turbulent vortices
(figures~\ref{fig:P-y}, \ref{fig:P_eps}\fa{a} and \ref{fig:epsp}\fa{a}). This
reduction of the energy production (figure~\ref{fig:P-y}) is caused by the
attenuation of turbulent vortices which are relevant to the production of the
Reynolds stress (figure~\ref{fig:U}\fa{b}); consequently, turbulent kinetic
energy is also attenuated. In contrast, the energy dissipation rate is not
significantly modulated in the examined cases (figure~\ref{fig:P_eps}\fa{b}).
Therefore, the energy production rate is described solely by the additional
energy dissipation rate $\epsilon_p$. Our DNS results
(figure~\ref{fig:Ar-epsp}\fa{a}) show that this energy balance, described by
(\ref{eq:ArP}), holds for the spatial averaging by numerically evaluating the
additional energy dissipation rate $\langle \epsilon_p \rangle$ (\ref{eq:epsp}).
We also show that we can describe the average attenuation rate $\mathcal{A}$
(\ref{eq:ArK}) of turbulent kinetic energy in terms of $\langle \epsilon_p
\rangle$ (figure~\ref{fig:Ar-epsp}\fa{c}). This attenuation mechanism well
explains the Stokes-number and particle-diameter dependence of $\langle K
\rangle$ (figure~\ref{fig:ArK}), since the energy dissipation rate $\langle
\epsilon_p \rangle$ due to particles becomes larger as (i) the Stokes number
becomes larger or (ii) the particle size becomes smaller under the condition
that the volume fraction is constant (figure~\ref{fig:epsp}\fa{b}).

To quantitatively predict the degree of turbulence attenuation, we estimate the
relative velocity required for the estimation (\ref{epsp_sh_ratio}) of the
additional energy dissipation rate $\langle \epsilon_p^\# \rangle/\langle
\epsilon_\times^\# \rangle$. Our DNS results (figure~\ref{fig:du}) demonstrate
that the relative velocity averaged in the outer layer is determined by the
functions (\ref{eq:bala}) of $St_h$. These are derived based on the argument by
\citet{Balachandar2009} for heavy pointwise particles, allowing us to quantify
$\langle \epsilon_p^\# \rangle/\langle \epsilon_\times^\# \rangle$ only from
particle properties through (\ref{eq:epsp_pre}). Moreover, our estimation of the
average turbulence attenuation in terms of $\langle \epsilon_p^\#
\rangle/\langle \epsilon_\times^\# \rangle$ well describes not only the present
DNS data but also previous experimental results by \citet{Kulick1994}
(figure~\ref{fig:Ar-epsp_sh}).

\section*{Acknowledgements}

This study was partly supported by the JSPS Grants-in-Aid for Scientific
Research 20H02068 and 23K13253. The DNS were conducted by using the
computational resources of the supercomputers Fugaku through the HPCI System
Research Projects (hp220232, hp230288 and hp240278). The numerical analyses were
conducted under the auspices of the NIFS Collaboration Research Program
(NIFS22KISS010 and NIFS24KISC007). We would like to thank Professor Uhlmann for
discussing the present study during our stay in Karlsruhe. We also thank
Professor Balachandar for discussing the evaluation of the relative velocity. 

\section*{Declaration of interests}
The authors report no conflict of interest.

\appendix

\section{Indicator proposed by \citet{Yu2021} for turbulence
enhancement\label{sec:enhancement}}

\begin{figure}
    \centering
    \includegraphics[width=90mm]{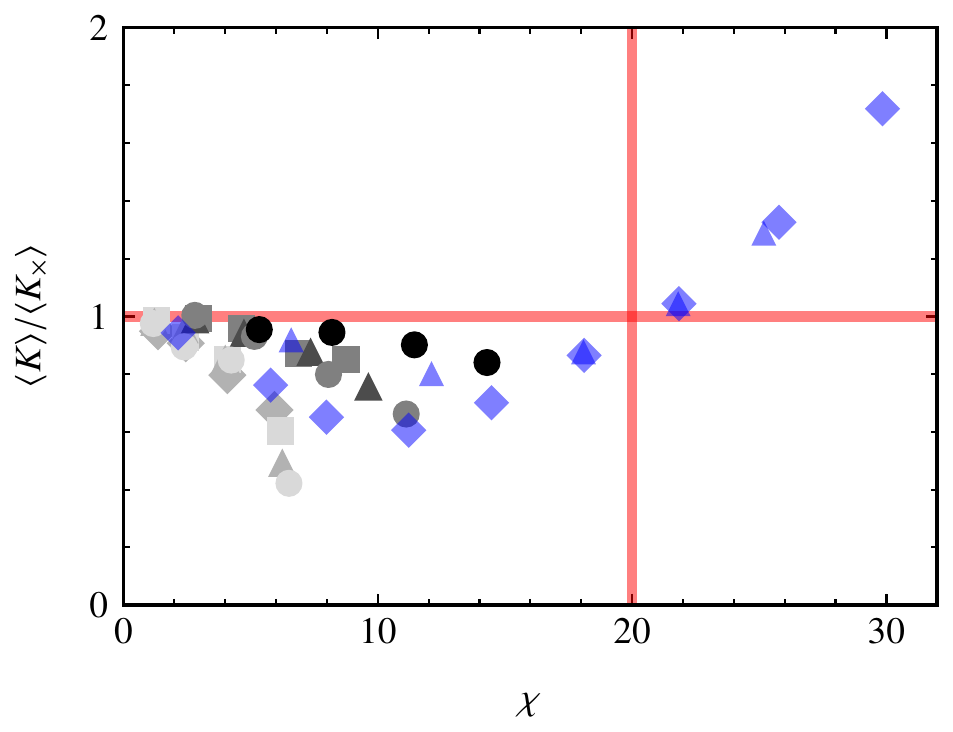}
    \caption{
	Spatial average $\langle K \rangle$ of turbulent kinetic energy
	normalized by the value $\langle K_\times \rangle$ in the single-phase flow
	as a function of the indicator $\chi$ defined as (\ref{eq:chi}). The grey
	symbols are the same as in figure~\ref{fig:ArK}. The blue symbols show the
	results by \citet{Yu2021} for particles with $D/h=0.1$, $\rho_p/\rho_f=2$
	and $\Lambda_0=0.0236$ but with different values of the settling velocity
	$u_g$ in turbulent channel flow at $Re_\tau \approx 350$ (triangles) and
	$180$ (diamonds).
    \label{fig:chi}
    }
\end{figure}

In the present study, turbulence enhancement does not occur because of the
absence of gravity. \citet{Yu2021} introduced the indicator defined as
\begin{align}
\chi
= 
\dfrac{Re_p}{Re_b^{0.33}\:(D/h)^{0.61}\:(\rho_p/\rho_f)^{0.05}} ,
\label{eq:chi}
\end{align}
by fitting their DNS results, and demonstrated that turbulence enhancement
occurs for $\chi \gtrsim 20$. This condition requires large $Re_p$, which is
also a known condition for turbulence enhancement \citep{Uhlmann2008,
Balachandar2010}. However, since the relative velocity in the present system
without gravity is in the order of the fluctuation fluid velocity at most
(figure~\ref{fig:du}), $Re_p$ is much smaller than $Re_b$. Therefore, the
condition $\chi \gtrsim 20$ is not satisfied; as shown with the grey symbols in
figure~\ref{fig:chi}. This is consistent with the fact that we observe only
attenuation in the present system. In contrast, the blue symbols show the
results by \citet{Yu2021} for small particles ($D^+ \approx 10$) at $Re_\tau
\approx 350$ (triangles) and $180$ (diamonds) in the presence of gravity. When
$\chi$ is larger than $20$ due to the gravitational effect, the average
turbulent kinetic energy is enhanced.

\section{Evaluation of energy dissipation rate\label{sec:eps}}

In this appendix, we describe how to accurately evaluate the energy
dissipation rate $\epsilon = 2 \nu \tensor{S}\mathbin{:}\tensor{S}$, where
$\tensor{S}$ is the strain-rate tensor, in DNS using the IBM. In each numerical
step of the IBM, we first ignore the presence of particles to obtain the
preliminary velocity $\vb*{u}^*$ by integrating the Navier--Stokes equation. We
then evaluate the body force to eliminate the velocity difference between
$\vb*{u}^*$ and the velocity $\vb*{u}_s$ on the particle surface. For particles,
the force $\vb*{F}^{IBM}_\ell$ acts on the Lagrangian points (the $\ell$th point
is denoted by $\vb*{x}_\ell$) distributed on the particle surface; whereas the
fluid is forced by $\vb*{f}^{IBM}_\ell = -\vb*{F}^{IBM}_\ell
\delta(\vb*{x}-\vb*{x}_\ell)$ on grid points $\vb*{x}$ around $\vb*{x}_\ell$,
where $\delta$ is a regularized delta function satisfying $\int_{\Omega_\ell} \:
\delta(\vb*{x}-\vb*{x}_\ell) \: \d V = 1$ with $\Omega_\ell$ being the spherical
region with the centre at $\vb*{x}_\ell$ and radius $1.5\Delta$. Note that the
sum of the rates of work due to $\vb*{F}^{IBM}_\ell$ and $\vb*{f}^{IBM}_\ell$ is
expressed by
\begin{align}
W^{IBM}_\ell + w^{IBM}_\ell
=
-
\int_{\Omega_\ell} 
\: 
\vb*{f}^{IBM}_\ell 
\cdot 
\qty(\vb*{u}_s(\vb*{x}_\ell) - \vb*{u}(\vb*{x}))
\:
\d V .
\label{eq:W+w}
\end{align}
Here, $W^{IBM}_\ell$ ($=\vb*{F}^{IBM}_\ell \cdot \vb*{u}_s$)
is the rate of the work due to $\vb*{F}^{IBM}_\ell$, whereas $w^{IBM}_\ell$ ($=
\int_{\Omega_\ell} \: \vb*{f}^{IBM}_\ell \cdot \vb*{u} \: \d V$) is rate of the
work due to $\vb*{f}^{IBM}_\ell$. The total rate of work (\ref{eq:W+w}) does
not vanish because of the velocity difference $\vb*{u}_s-\vb*{u}$,
which results in the extra energy dissipation, as numerically demonstrated
below. Therefore, in the present analyses, we take into account this
contribution due to the IBM, and we evaluate turbulent energy dissipation rate
as
\begin{align}
\epsilon
=
2 \nu \tensor{S} \mathbin{:} \tensor{S}
+
\sum_{\ell}
\:
\vb*{f}^{{IBM}}_\ell \cdot \qty(\vb*{u}_s(\vb*{x}_\ell) -
\vb*{u}(\vb*{x})) . 
\label{eq:eps}
\end{align}
Here, we use the value of $\tensor{S}$ evaluated in the DNS, and $\sum_{\ell}$
denotes the summation over the Lagrangian points $\vb*{x}_\ell$ located within
the distance $1.5\Delta$ from $\vb*{x}$.

\begin{figure}
	\centering
	\includegraphics[width=90mm]{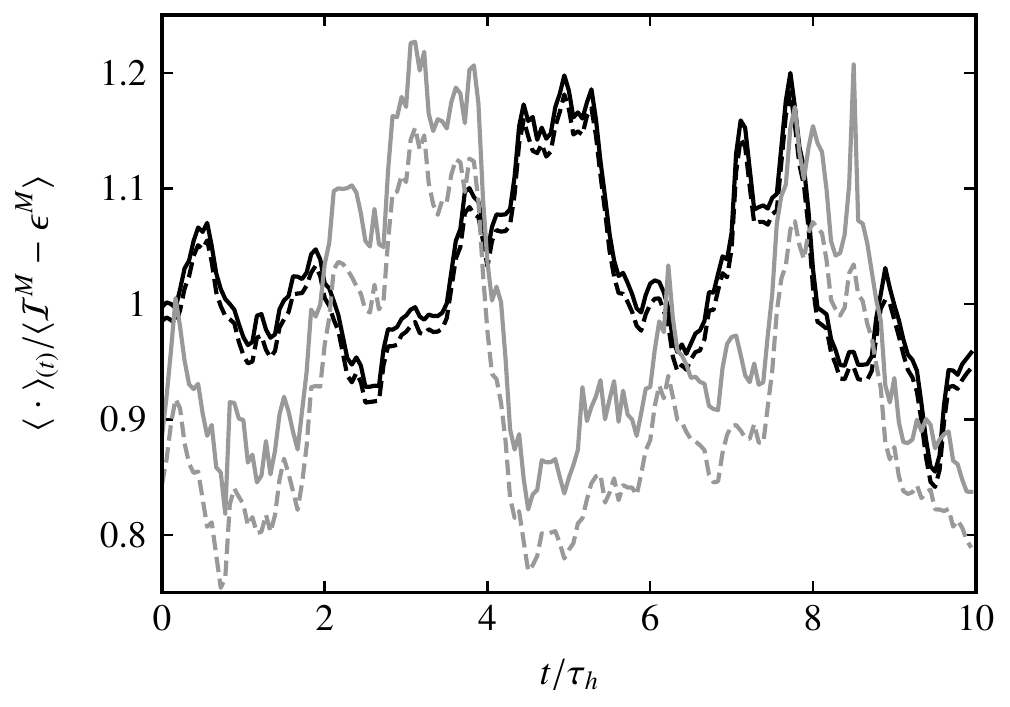}
	\caption{Spatial average $\langle \: \cdot \: \rangle_{(t)}$ at 
	time $t$ for turbulent energy dissipation rate $\epsilon$ (solid line)
	estimated by (\ref{eq:eps}) and that for $\epsilon^*$ (dashed one) in the
	case ignoring the second term of (\ref{eq:eps}). The black and grey lines
	show the results for $St_h=0.16$ and $10$, respectively, for turbulence at
	$Re_\tau=180$. The values are normalized by $\langle \mathcal{I}^M -
	\epsilon^M \rangle$, which is balanced with the time average $\langle
	\epsilon \rangle$ of $\langle \epsilon \rangle_{(t)}$. 
	\label{fig:eps_fdu}}
\end{figure}

\begin{figure}
	\centering
	\includegraphics[width=105mm]{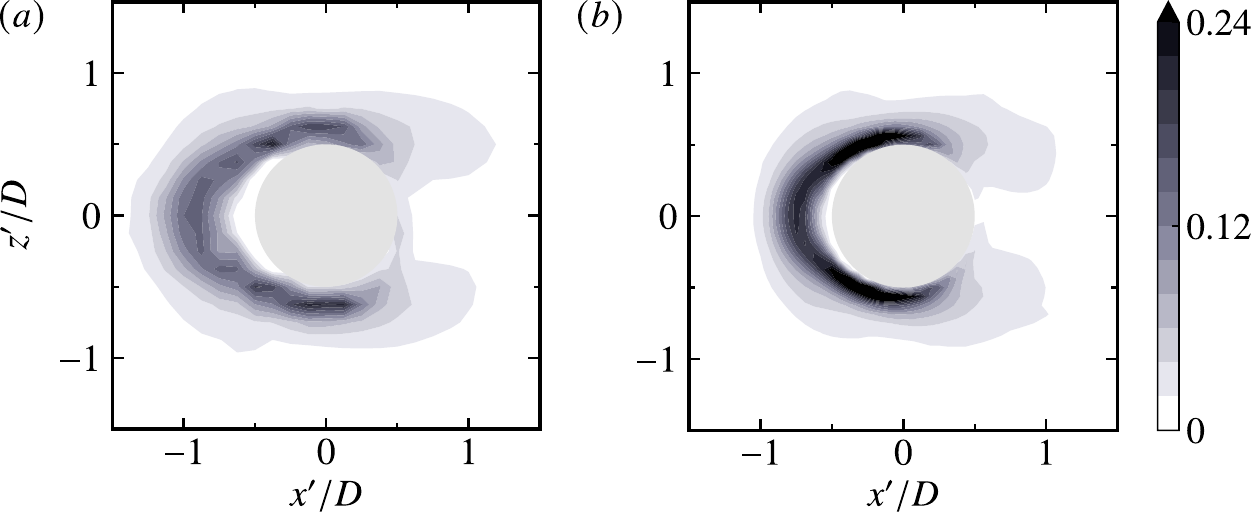}
	\caption{
	Average spatial distribution of turbulent energy dissipation rate $\epsilon$
	around a particle with $D/h=0.089$ and $St_h=10$ located at the channel
	centre in turbulence at $Re_\tau=180$. The results are obtained using the
	different DNS resolutions (\fa{a}) $D/\Delta =8$ and (\fa{b}) $16$. The
	values are shown in wall units. 
    \label{fig:eps_ar}
	}
\end{figure}

To verify the above argument, we show the spatial average $\langle \: \cdot \:
\rangle_{(t)}$ of $\epsilon$ at each time $t$ by the solid line in
figure~\ref{fig:eps_fdu}. The dashed line is the average $\langle \epsilon^*
\rangle_{(t)}$ ($= \langle 2 \nu \tensor{S} \mathbin{:} \tensor{S}
\rangle_{(t)}$) in the case ignoring the second term of (\ref{eq:eps}). The
black and grey lines show the results for $St_h=0.16$ and $10$, respectively. We
see that the discrepancy between the solid and dashed lines is non-negligible
especially for larger $St_h$. Note that the solid line always lies above the
dashed one, which means that the second term of (\ref{eq:eps}) contributes to
energy dissipation. We explain this by rewriting the summand as 
\begin{align}
\vb*{f}^{IBM}_\ell \cdot \qty(\vb*{u}_s - \vb*{u})
\approx
\dfrac{ (\vb*{u}_s - \vb*{u})^2 }{
\Delta t } \: \delta(\vb*{x}-\vb*{x}_\ell) 
>
0
\end{align}
because $\vb*{F}^{IBM}_\ell = -{(\vb*{u}_s - \vb*{u}^*)}/{ \Delta t }$ and
$\vb*{u}^*$ is close to $\vb*{u}$. We also verify (\ref{eq:eps}) by the energy
budget that the temporal average $\langle \epsilon \rangle$ of $\langle \epsilon
\rangle_{(t)}$ is balanced with the average energy input rate $\langle
\mathcal{I}^M \rangle$ minus the average energy dissipation rate $\langle
\epsilon^M \rangle$ in the mean flow. We have confirmed that the balance is
satisfied with an error within $1$\:\%, and if we ignored the second term in
(\ref{eq:eps}), the average value $\langle \epsilon^* \rangle$ would give a
discrepancy of $8$\:\% to $\langle \mathcal{I}^M - \epsilon^M \rangle$ for
$St_h=10$. We therefore evaluate $\epsilon$ using (\ref{eq:eps}) in the present
study.

Thus, the contribution of the second term of (\ref{eq:eps}) to energy
dissipation rate is non-negligible. However, this does not mean that the present
DNS is not enough to resolve energy dissipation rate field around particles,
although the second term vanishes in the limit of infinite resolution.
Figure~\ref{fig:eps_ar} shows the average distribution of energy dissipation
rate $\epsilon$ on the two-dimensional plane $(x^\prime, z^\prime)$ that passes
through the particle centre. The results are for particles with $D/h = 0.089$
and $St_h= 10$ existing at the channel centre in turbulence at $Re_\tau=180$.
Panel (\fa{a}) shows the case with the resolution ($D/\Delta =8$) in the present
DNS; while panel (\fa{b}) shows the case with twice the resolution ($D/\Delta
=16$). We see that the higher resolution captures stronger $\epsilon$ near the
particle interface, as was also observed by \citet{Xia2022}. It is however
important to note that since the energy dissipation rate is more widely
distributed for the lower resolution, average values are almost the same; more
specifically, (\fa{a}) $0.20$ and (\fa{b}) $0.21$ in wall units. Therefore, the
present resolution is sufficient to estimate the average energy dissipation rate
$\epsilon_p$ around particles, which is a key quantity for average attenuation
rate of turbulence.

\section{Mean velocity and Reynolds stress\label{sec:U}}

\begin{figure}
	\centering
	\includegraphics[width=\textwidth]{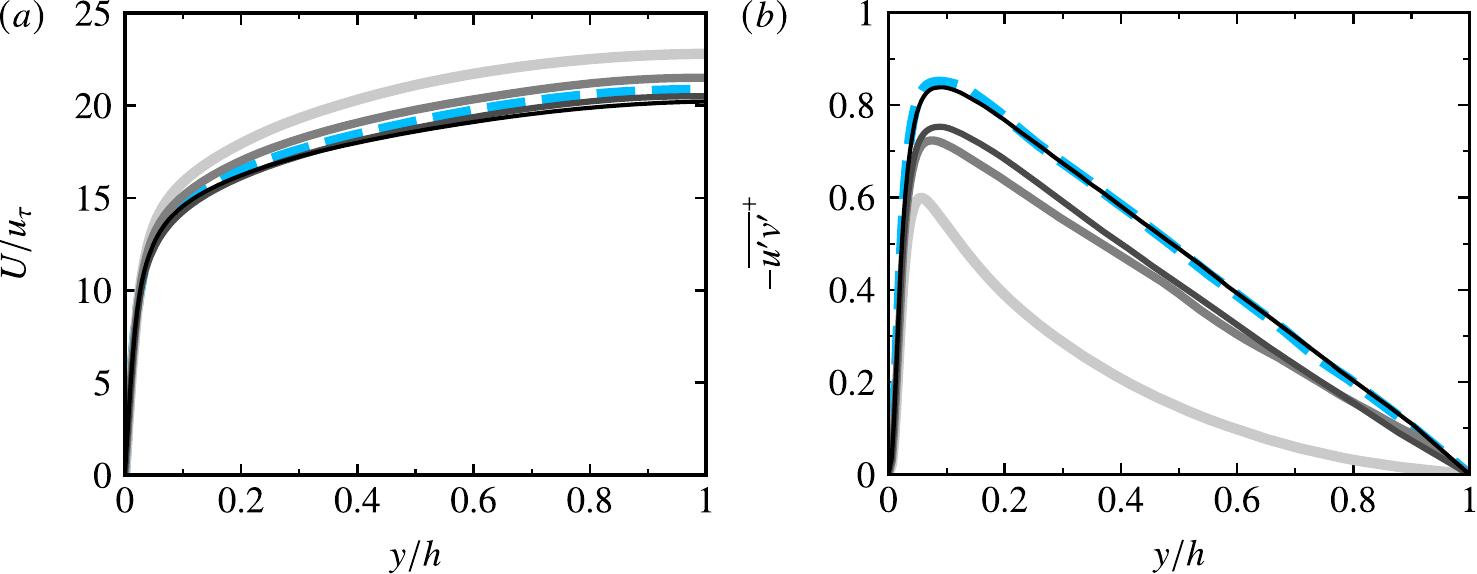}
	\caption{
	Wall-normal profiles of (\fa{a}) the mean streamwise velocity $U(y)$ and
	(\fa{b}) Reynolds stress $-\overline{u^\prime v^\prime}(y)$ at
	$Re_\tau=512$. The lines indicate the same parameters as shown in
	figures~\ref{fig:K-y}(\fa{a}) and \ref{fig:P-y}(\fa{a}). 
    \label{fig:U}}
\end{figure}

We have shown in figure~\ref{fig:P-y} the significant reduction of energy
production rate $\mathcal{P}^{t \leftarrow M}$ from the mean flow. In this
appendix, to show that this reduction is due to the attenuation of the Reynolds
stress, we show in figure~\ref{fig:U} the wall-normal profiles of (\fa{a}) the
mean streamwise velocity $U$ and (\fa{b}) Reynolds stress $-\overline{u^\prime
v^\prime}$ at $Re_\tau=512$. Particle parameters are the same as in
figure~\ref{fig:P-y}(\fa{a}); namely, the results for the common $D/h$
($=0.031$) but different values of $St_h$. Looking at
figure~\ref{fig:U}(\fa{a}), we notice that the mean velocity is not drastically
modulated from the single-phase flow (blue dashed line). More precisely, the
mean bulk velocity slightly increases with $St_h$, which is related with the
reduction of the Reynolds stress. This is verified by
figure~\ref{fig:U}(\fa{b}), where we see that the Reynolds stress is attenuated
more significantly as $St_h$ gets larger. 

\section{Energy input rate and dissipation rate in the mean flow\label{sec:MKE}}

\begin{figure}
	\centering
	\includegraphics[width=1.0\textwidth]{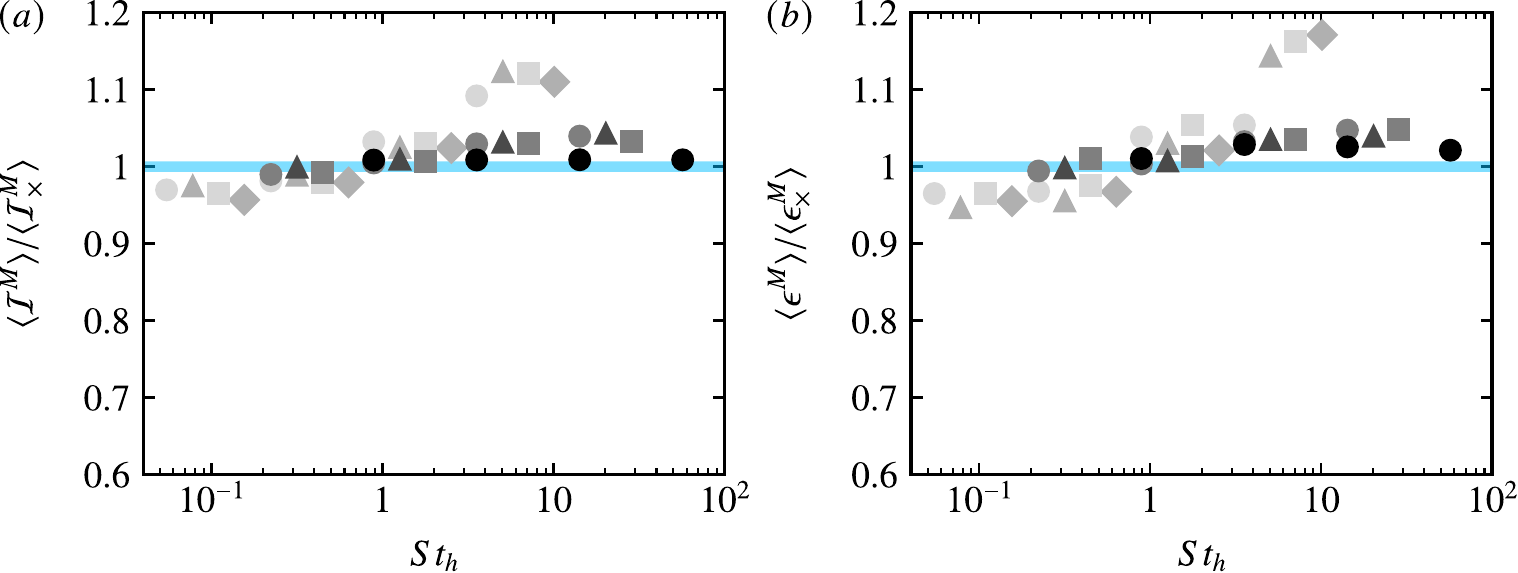}
	\caption{
	Spatial average of (\fa{a}) energy input rate $\mathcal{I}^M$ due to the
	external pressure gradient and (\fa{b}) energy dissipation rate $\epsilon^M$
	in the mean flow. The values are normalised by those in the single-phase
	flow. The symbols are the same as in figure~\ref{fig:ArK}. 
    \label{fig:MKE}
	}
\end{figure}

Figure~\ref{fig:MKE} shows the averages of (\fa{a}) energy input rate
$\mathcal{I}^M$ ($=U f_x^{ex}$) due to the external pressure gradient and
(\fa{b}) energy dissipation rate $\epsilon^M$ ($=\nu ({\d U}/{\d y})^2$) in the
mean flow. These quantities show similar monotonic trends for $St_h$ due to the
slight modulation of the mean velocity (figure~\ref{fig:U}\fa{a}). As a result,
the difference between these quantities, that is the turbulent dissipation rate
$\langle \epsilon \rangle$ ($= \langle \mathcal{I}^M \rangle - \langle
\epsilon^M \rangle$), is only slightly modulated (figure~\ref{fig:P_eps}\fa{b}). 

\section{Drag reduction\label{sec:Cf}}

\begin{figure}
	\centering
	\includegraphics[width=90mm]{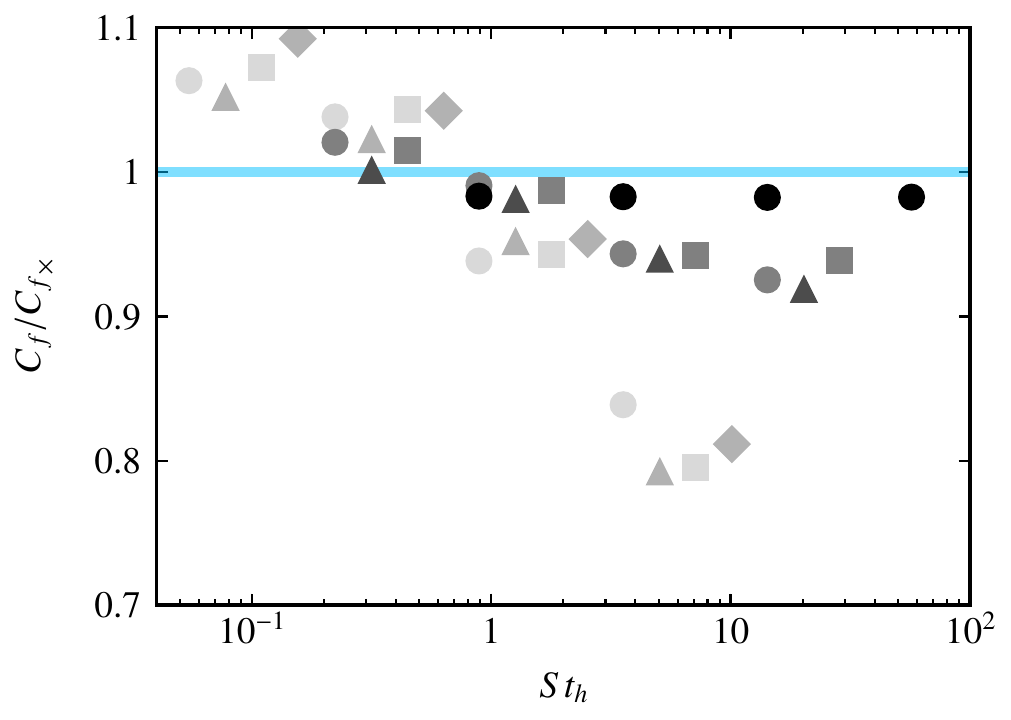}
	\caption{Frictional drag coefficient $C_f$ normalized by the value
	in the single-phase flow. The symbols are the same as in
	figure~\ref{fig:ArK}. 
	\label{fig:Cf}}
\end{figure}

We show in figure~\ref{fig:Cf} the friction drag coefficient $C_f$ normalized by
the value in the single-phase flow as a function of $St_h$, where $C_f = \tau_w
/ (\rho_f U_b^2/2) = 2 u_\tau^2 / U_b^2$ with $\tau_w$ being the average
frictional stress on the walls. Note that in the present study, the external
pressure gradient to drive the flow, and therefore the frictional stress
$\tau_w$ are constant; while the mean bulk velocity $U_b$ can be modulated due
to the presence of particles. Therefore, larger (or smaller) $U_b$ means smaller
(or larger) $C_f$. We see in figure~\ref{fig:Cf} that for $St_h \gtrsim 1$, the
ratio $C_f / {C_f}_\times$ is less than $1$. More concretely, $C_f$ gets smaller
for larger $St_h$ and smaller $D/h$. This is similar to the behaviour of average
of turbulent kinetic energy (figure~\ref{fig:ArK}) and Reynolds stress
(figure~\ref{fig:U}\fa{b}). We therefore conclude that when turbulence is
attenuated, the frictional drag coefficient is also reduced. Drag modulations were
also investigated in turbulent channel flow laden with spheroidal particles
\citep{Ardekani2017, Ardekani2019}, neutrally buoyant spheres \citep{Picano2015}
and small heavy ones \citep{Yu2017, Costa2021, Dave2023}. 

\section{Wall-normal profile of the relative velocity\label{sec:du-y}}

\begin{figure}
	\centering
	\includegraphics[width=\textwidth]{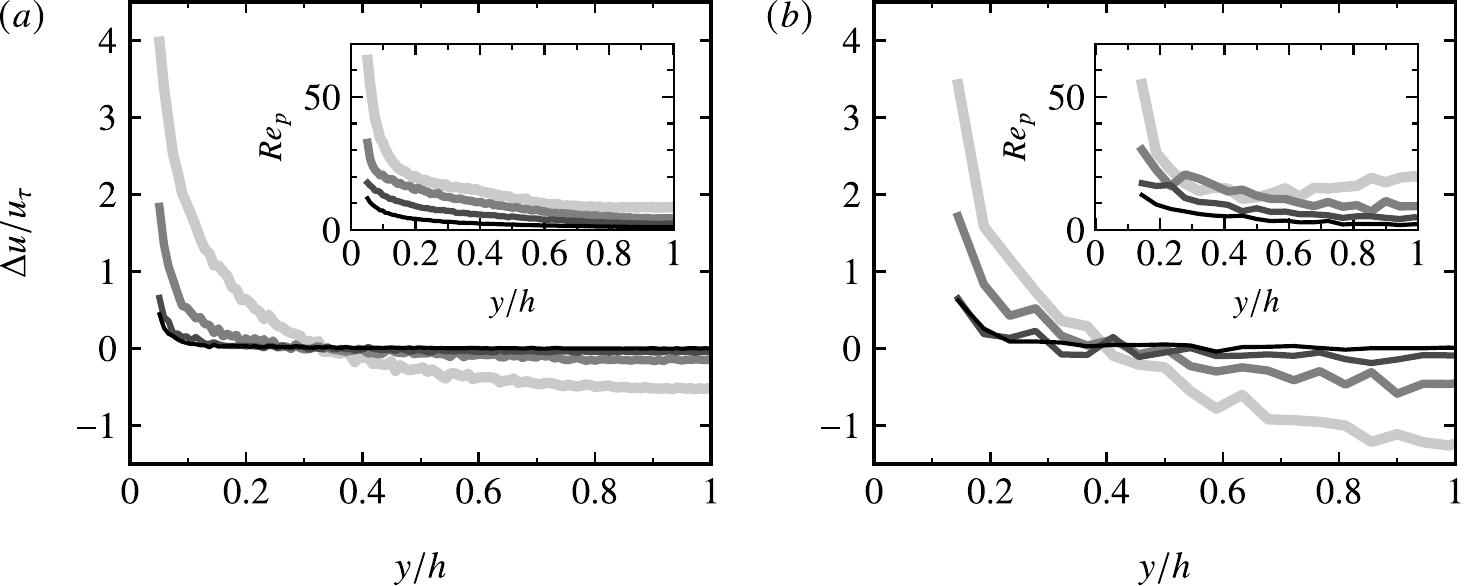}
	\caption{
	Wall-normal profiles of the mean relative velocity $\Delta u(y)$ for
	$D^+=16$ [i.e.~(\fa{a}) $D/h=0.031$ and (\fa{b}) $0.089$] in turbulence at
	(\fa{a}) $Re_\tau=512$ and (\fa{b}) $180$. The four lines in each panel
	show, from the thinner (and darker) to thicker (and lighter), (\fa{a})
	$St_h=0.056$, $0.22$, $0.89$ and $3.6$, and (\fa{b}) $St_h=0.16$, $0.63$,
	$2.5$ and $10$. The insets show the particle Reynolds number $Re_p(y)$.
	\label{fig:du-y}}
\end{figure}

As mentioned in \S\:\ref{sec:du-outer}, the relative velocity depends strongly
on height. We can confirm this in figure~\ref{fig:du-y}, which shows the
wall-normal profiles of the mean streamwise velocity difference $\Delta u (y)$
for (\fa{a}) $Re_\tau=512$ and (\fa{b}) $180$. These results are for common
$D/h$ at each $Re_\tau$ but for different $St_h$. The inset shows the particle
Reynolds number $Re_p(y) = |\Delta u(y)| D/\nu$. We see that when $St_h$ is
larger, $|\Delta u|$, and therefore $Re_p$, are also larger. In the case of
$St_h \gtrsim 1$, we also see that the sign of $\Delta u$ changes around $y/h
\approx 0.3$ irrespective of $Re_\tau$. Since particles with $St_h \gtrsim 1$
cannot follow the swirls of outer-layer vortices, they are slower than fluid in
the outer layer. On the other hand, when the particles are swept out by the
outer-layer vortices and move into the slower inner layer, the particles become
faster than the fluid. Thus, the relative velocity behaves in a qualitatively
different manner between the inner and outer layers. In \S\:\ref{sec:du-outer},
we have defined the outer layer as $y/h \geq 0.3$, where $\Delta u$ depends
weakly on $y$ and $\Delta u > 0$. 

\end{document}